\documentclass[aps,prb,amsmath,amssymb,superscriptaddress,floatfix,showpacs,twocolumn]{revtex4-1}

\usepackage{graphicx}
\usepackage{mathdots}
\usepackage{epstopdf} 
\usepackage[usenames,dvipsnames]{color}
\usepackage{bm}

\usepackage{inputenc}
\usepackage{xcolor}
\usepackage{tikz}
\usepackage{physics}

\usepackage{booktabs}

\usepackage[colorlinks,bookmarks=false,citecolor=blue,linkcolor=red,urlcolor=blue]{hyperref}
\usepackage[colorlinks,bookmarks=false,citecolor=blue,linkcolor=red,urlcolor=blue]{hyperref}
\usepackage{xcolor}
\usepackage{mathdots}
\usepackage{float}
\usepackage{needspace}
\usepackage{soul}
\usepackage[usenames,dvipsnames]{color}

\newcommand{\noteZ}[1]{{\color{blue} \emph{ GyZ: #1}}}

\begin{document}

\preprint{APS/123-QED}

\title{Electrically driven spin resonance with bichromatic driving}

\author{Zolt\'an Gy\"orgy}
\affiliation{Institute of Physics, 
E\"otv\"os University, 
H-1117 Budapest, Hungary}

\author{Andr\'as P\'alyi}
\affiliation{Department of Theoretical Physics, Institute of Physics, Budapest University of Technology and Economics, M\H{u}egyetem rkp. 3., H-1111 Budapest, Hungary}
\affiliation{MTA-BME Quantum Dynamics and Correlations Research Group, Budapest University of Technology and Economics, Műegyetem rkp. 3., H-1111 Budapest, Hungary}

\author{G\'abor Sz\'echenyi}
\affiliation{Institute of Physics, 
E\"otv\"os University, 
H-1117 Budapest, Hungary}

\date{\today}

\begin{abstract}
Electrically driven spin resonance (EDSR) is an established tool for controlling semiconductor spin qubits. Here, we theoretically study a frequency-mixing variant of EDSR, where two driving tones with different drive frequencies are applied, and the resonance condition connects the spin Larmor frequency with the \emph{sum} of the two drive frequencies. Focusing on flopping-mode operation of  a  single electron in a double quantum dot with spin-orbit interaction, we calculate the parameter dependence of the Rabi frequency and the Bloch-Siegert shift. A shared-control spin qubit architecture could benefit from this bichromatic EDSR scheme, as it enables simultaneous single-qubit gates.
\end{abstract}

\maketitle

\section{Introduction}

The spin degree of freedom of an electron confined in a quantum dot naturally defines a qubit \cite{PhysRevA.57.120}. Spin qubits in semiconductors \cite{burkard2021semiconductor}, such as germanium \cite{Scappucci2021}  or silicon \cite{RevModPhys.85.961}, are promising candidates for the building block of a future scalable fault-tolerant  quantum computer due to the long qubit lifetimes, the high gate fidelities beyond the error-correction threshold  \cite{mills2021twoqubit,Noiri2022,Xue2022}, and the small footprint. In spite of these advantages, in terms of the number of qubits, the state-of-the-art spin qubit quantum  processor \cite{Hendrickx2021, Philips2022, Madzik2022} lags behind  superconducting  and trapped-ion-based  quantum processors.

A conventional way of realizing a single-qubit gate for a spin qubit is electron spin resonance (ESR). There, the electron is illuminated by an ac magnetic field \cite{PhysRevLett.100.236802}. If the frequency of the field $\omega$ is matching with the qubit splitting $\omega_\textrm{split}$, then the spin performs coherent Rabi oscillation. It is technically demanding to selectively address qubits with magnetic field at nanoscale, therefore it is more convenient to replace the magnetic driving by electric dipole spin resonance (EDSR) technique, where the electric field couples to the spin via spin-orbit coupling \cite{doi:10.1126/science.1148092}, hyperfine interaction \cite{PhysRevLett.99.246601} or a magnetic-field gradient\cite{PhysRevLett.96.047202}. Using an ac electric field also simplifies the device:  instead of extra microwave antennas, EDSR is triggered by modulating the gate voltages of the quantum dots. EDSR can be further enhanced in the flopping-mode configuration, i.e., when a single electron in a double quantum dot is tuned to the charge tipping point \cite{PhysRevB.100.125430, PhysRevResearch.2.012006}.

Rabi oscillation occurs not only at the fundamental resonance,  $\omega=\omega_\textrm{split}$, but also when the  driving field is a subharmonic of the qubit splitting,  $\omega=\omega_\textrm{split}/N$ with $N\in\mathbb Z^+$. Subharmonic Rabi oscillation\cite{PhysRevLett.115.106802, PhysRevB.100.214205} has been widely investigated in the literature and has been observed experimentally in electrically driven quantum dots \cite{Pei2012, PhysRevB.89.115409, Laird_2009, https://doi.org/10.48550/arxiv.2205.04905}. Besides subharmonic excitation, two-photon spin transition can also be realized by \emph{bichromatic} excitation, i.e., by multiplexing two distinct ac fields with frequencies $\omega$ and $\bar\omega$. This bichromatic excitation induces Rabi oscillation if the frequencies satisfy the resonance condition  $\omega+\bar\omega=\omega_\textrm{split}$. In this case, the qubit transition is caused by the interplay of the two ac fields. 

Bichromatic driving is a widely used experimental technique \cite{PhysRevResearch.3.033004, doi1808697, ELES2010232}, in particular, when orthogonal radio frequency and microwave  magnetic fields are applied  \cite{jeschke1999coherent,fedoruk2004transient,saiko2008effect,fedoruk2009pulsed}.
%in order to the reduce the adverse effect of inhomogenous microwave field.\cite{saiko2018suppression} 
In this paper, we focus on bichromatic driving such that the two frequencies are of the same order of magnitude, similar to the experiment in Ref. \onlinecite{PhysRevB.92.245422}, where Landau-Zener transition in a double quantum dot was induced by bichromatic electric driving.

\begin{figure}[t]
\centering
\includegraphics[width=0.9\columnwidth]{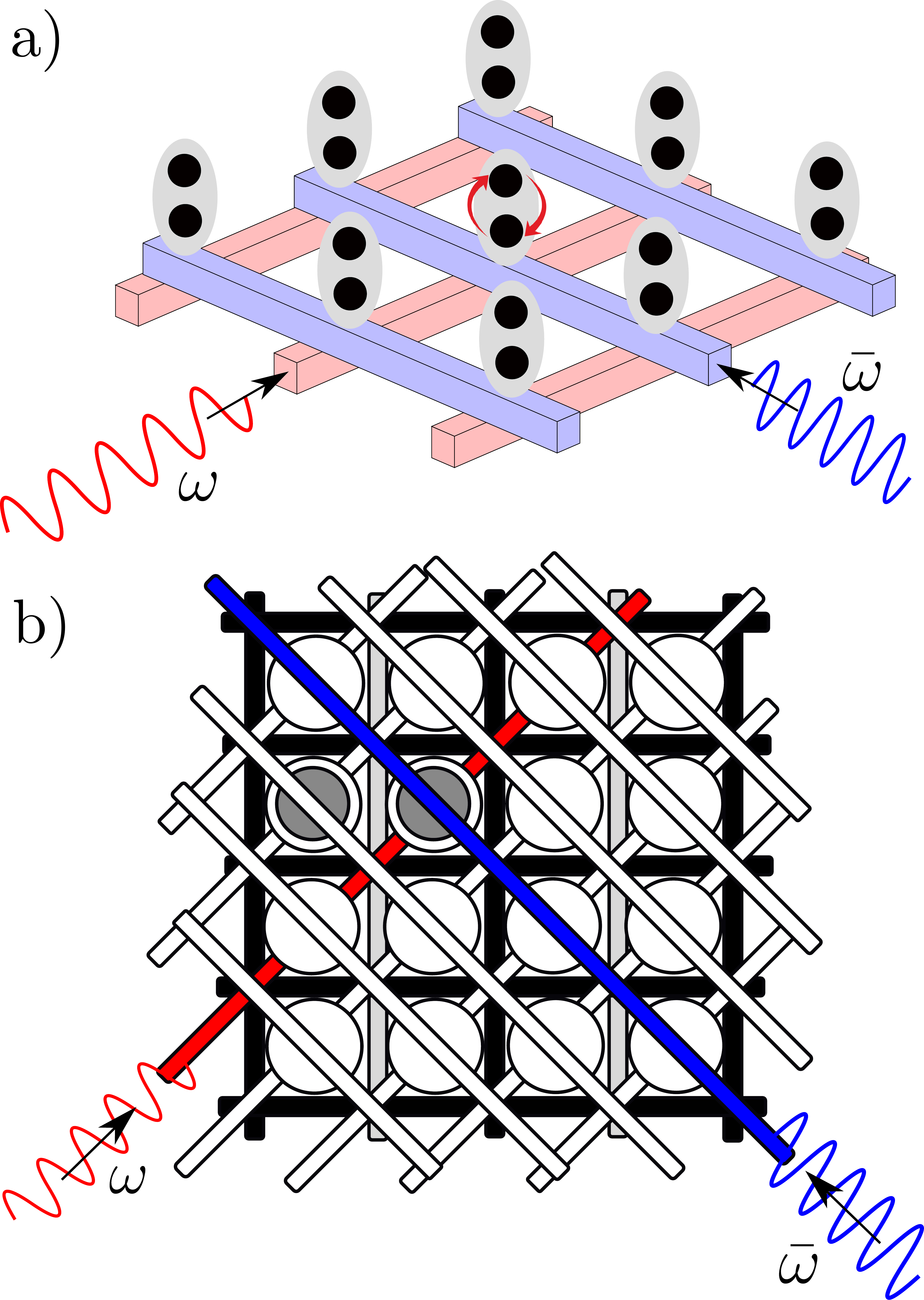}

\caption {\label{fig:setup} Bichromatic driving used for selective qubit addressing in 2D qubit array. 
(a)  Vertical arrangement, inspired by Ref.~\onlinecite{doi:10.1126/sciadv.1500707}. Qubits (grey ellipses) are defined by the charge or spin of single electrons confined in double quantum dots (black dots). Selective qubit addressing is realized by bichromatic driving, where ac voltages with frequency $\omega$ and $\bar\omega$ are applied on two orthogonal plunger gates (blue and red lines). If $\omega+\bar\omega$ matches the qubit splitting, then only the qubit at the intersection of the two driven gates (i.e., the central qubit in the figure) performs Rabi oscillations.
Electron confinement might require further gates that are not drawn here.
%For simplicity, the gate structure creating the confinement potential for the double quantum dots is not drawn.
(b) Lateral arrangement, inspired by Ref.~\onlinecite{doi:10.1126/sciadv.aar3960}. When performing single-qubit gates, the grey barrier gates allow hybridization between quantum dots, while the black ones do not. Qubits are defined by the charge or spin of electrons confined in a double quantum dot. The qubit addressed by bichromatic driving is represented by the two grey disks. On-site energy is controlled by the diagonal plunger gates. Selective addressing of the grey qubit is realized by bichromatic driving, where ac voltages with frequency $\omega$ and $\bar\omega$ are switched on two orthogonal plunger gates (red and blue lines). If $\omega+\bar\omega$ matches the qubit splitting, then only the selected (grey) qubit is controlled.}
\end{figure}

What are the advantages and disadvantages of bichromatic driving? A disadvantage is a reduced Rabi frequency, as a bichromatic transition is a second-order effect in the drive strengths. An advantage, as we argue below, is that it opens up new opportunities for selective addressing and gate parallelization in a 2D qubit array with shared control. Another potential advantage is that bichromatic control requires lower driving frequencies compared to the fundamental resonance. This can be a beneficial feature, as in common experimental setups, the attenuation of the drive signal in the GHz range increases with increasing frequency.

To expose these opportunities, we envision and describe two setups, as shown in Fig.~\ref{fig:setup}, both hosting a 2D quantum-dot array. The first setup, see Fig.~\ref{fig:setup}a, features vertically arranged double quantum dots (black disks), each holding a single electron. This setup is inspired by the architecture proposed in  Ref.~\onlinecite{doi:10.1126/sciadv.1500707}. Qubits (grey ellipses) are defined by the charge or spin degree of freedom of the electrons, and their high degree of uniformity is assumed. 
%Similarly to the crossbar architecture introduced  in Ref.~\onlinecite{doi:10.1126/sciadv.aar3960} 
Two sets of long and straight plunger gates (red and blue) control the on-site energies of the quantum dots.
Red and blue plunger gates are perpendicular to each other, and qubits are located above the intersection points of the plunger gates. (These are only quasi-intersection points, because the plunger gates do not touch each other.)
Note that each plunger gate controls the on-site energies of the line of quantum dots located above the gate. 
Here, electric-field-induced selective addressing of a single qubit can be realized by switching on  ac voltages with frequency $\omega$ and $\bar\omega$ on two orthogonal plunger gates. If the sum of the two frequencies matches the qubit splitting, then, because the electric field is well localized, only the selected qubit above the intersection point of the plunger gates is addressed, as shown in Fig.  \ref{fig:setup}(a). 

Another setup, featuring a lateral arrangement of double quantum dots, motivated by the crossbar architecture of Ref. \onlinecite{doi:10.1126/sciadv.aar3960}, is shown in Fig.~ \ref{fig:setup}(b). Both proposals will be discussed below, in Sec.~\ref{bichromatic_application}.

In both shared-control schemes outlined above and depicted in Fig.~\ref{fig:setup}, parallelization of single-qubit gates on different qubits is a challenge. We show here that multiplexing different frequencies on the plunger gates opens the possibility to realize different single-qubit gates on every qubit in a parallel way. A similar multiplexing technique was used recently for parallel dispersive  readout in a quantum dot array \cite{Ruffino2022}.  

The rest of the paper is arranged as follows. In Sec.~\ref{Bichromatic_ESR} and in the Appendix,  we  review the many-mode Floquet theory, and apply it to calculate the resonance condition and the Rabi frequency for magnetic-field-induced bichromatic electron spin resonance. Then, two sections are devoted to present our analytical and numerical results for electric-field-induced bichromatic driving of an electron defined in a double quantum dot. In Sec.~\ref{bichromatic_charge} the control of a charge qubit,  while in Sec.~\ref{bichromatic_EDSR}, the control of a spin qubit is discussed. In Sec.~\ref{bichromatic_application}, we analyze 2D qubit arrays in more detail, and as a main result, we show how bichromatic driving enables selective qubit addressing and quantum-gate parallelization. In Sec.~\ref{discussion} we summarize our results.

\section{Bichromatic electron spin resonance} \label{Bichromatic_ESR}

ESR is an established method to coherently control the spin quantum state of an electron. In the presence of an external magnetic field with strength $B$, the energy of the two spin states are split by $\omega_\textrm{split}=g\mu_BB$, where $g$ is the g factor and $\mu_B$ is the Bohr magneton. In case of monochromatic driving, a perpendicular ac magnetic field  with amplitude $B_\textrm{ac}$ and driving frequency $\omega$  induces coherent Rabi oscillation in the weak-driving limit $B \gg B_\textrm{ac}$, if the resonance condition $g\mu_BB=\hbar\omega$ is fulfilled.   In that case, the spin coherently evolves with the Rabi frequency $\Omega_\textrm{Rabi, mono}=\frac{g\mu_B B_\textrm{ac}}{2\hbar}$. 

The above description is based on the rotating-wave approximation, which is a standard technique to satisfactorily describe Rabi oscillations in ESR. However, fine details of the dynamics in the weak-driving limit, and pronounced effects in case of stronger driving, such as  the Bloch-Siegert shift, the Bloch-Siegert oscillations, and subharmonic resonances \cite{PhysRevB.92.054422} are not captured by this approximation; for those cases, Floquet theory is often used instead \cite{PhysRev.138.B979}.  
Floquet technique reformulates the spin dynamics in terms of the eigenvalue problem of an infinite, time-independent matrix. 
Furthermore, when two degenerate Floquet levels are weakly coupled to the other Floquet levels, a Schrieffer-Wolff transformation yields a $2\times 2$ matrix, allowing to read off the Rabi frequency and the resonance condition. 

%In this paper we need to  go beyond this approximation and use an analytical technique based on Floquet-theory. The single-mode Floquet-theory was introduced in Ref. \onlinecite{PhysRev.138.B979}, and the point of this technique is reformulating the Schrödinger-equation of the spin dynamics as an eigenproblem of an infinite matrix. In the weak driving limit two-fold degenerated Floquet-levels of such infinite matrix are weakly coupled to the other levels, therefore applying Schieffer-Wolff transformation we can derive an effective 2$\times$2 matrix whereform we can read out the Rabi frequency and the resonance condition. This technique was used not only for the fundamental but also for the subharmonic resonances.\cite{PhysRevB.92.054422}

In bichromatic ESR, two ac magnetic field components are applied, characterized by their drive strengths $B_\textrm{ac}$ and $\bar B_\textrm{ac}$, and the corresponding drive frequencies $\omega$ and $\bar\omega$. This system can be handled by the so-called many-mode Floquet theory \cite{ho1983semiclassical}, which is the generalization of single-mode Floquet theory in the presence of multiple periodic fields. This technique was used for a bichromatic ESR \cite{ho1984semiclassical}, where both ac fields are orthogonal to the static field, but in such case only an odd number of photons could be involved in the transition. To realize two-photon bichromatic ESR one ac field must have a component that is longitudinal (parallel) to the static field. This phenomenon was investigated theoretically \cite{saiko2010multiplication,saiko2015multi} when the bichromatic driving consists of a longitudinal radio-frequency field and a transverse microwave field.

In this section, we derive a formula for the Rabi frequency and a formula of the resonance condition, the latter incorporating the Bloch-Siegert shift. We focus on a special case of bichromatic ESR, when the $B_\textrm{ac}$ field is longitudinal (parallel), and  $\bar B_\textrm{ac}$ field is orthogonal, to the static magnetic field, furthermore, both frequencies are of the same order of magnitude. The corresponding Hamiltonian reads
\begin{eqnarray}\label{ESR_Hamiltonian}
   H(t)&=&\dfrac{1}{2}g\mu_BB\sigma_z+\dfrac{1}{2}g\mu_BB_{\mathrm{ac}}\cos{(\omega t)}\sigma_z  \nonumber\\&+&\dfrac{1}{2}g\mu_B\bar B_{\mathrm{ac}}\cos{(\bar\omega t)}\sigma_x,
\end{eqnarray}
where $\sigma_x$, $\sigma_z$ are Pauli matrices in the spin subspace. 

First, we apply many-mode Floquet theory in presence of two ac magnetic fields to find the solution of this time-dependent Hamiltonian. As it is presented in details in Appendix \ref{app_multi_floquet_theory}, this theory leads to an eigensystem problem of an infinite, so-called Floquet matrix.
It is instructive to regard the Floquet matrix of the undriven system as the unperturbed case, and consider the contribution of the driving fields to the Floquet matrix as a perturbation.
If $\omega + \bar\omega$ matches the qubit splitting, then the unperturbed Floquet matrix has double-degenerate eigenstates.
Next step, we choose one of these doublets, and restrict our analysis to the weak-driving regime, where $\left\{B_\textrm{ac},\bar B_\textrm{ac}\right\} \ll B$.
The two highlighted unperturbed Floquet levels are only weakly coupled to the rest of the matrix by the driving terms, hence an effective $2\times2$ Hamiltonian can be derived by a second-order Schrieffer-Wolff transformation. 

The resulting effective Hamiltonian describes complete Rabi oscillations with frequency
\begin{equation}\label{ESR_Rabi}
    \Omega_\textrm{Rabi,bi}=\dfrac{\left(g\mu_B\right)^2B_{\mathrm{ac}}\bar B_{\mathrm{ac}}}{4\hbar^2\omega},
\end{equation}
if the resonance condition is fulfilled
\begin{equation}\label{ESR_resonance_condition}
  \omega+\bar\omega=\omega_{\mathrm{split}}+\omega_{\mathrm{BS}},
\end{equation}
where 
\begin{equation}\label{ESR_BS}
   \omega_{\mathrm{BS}}=\dfrac{\omega_\textrm{split}\left(g\mu_B\bar B_{\mathrm{ac}}\right)^2}{4\hbar^2\omega\left(2\omega_\textrm{split}-\omega\right)} .
\end{equation}
For the detailed calculation, follow Appendix~\ref{app_Schrieffer_Wolff}. These formulas are valid, if the frequencies and amplitudes are of the same order of magnitude: $\omega \sim \bar\omega$, $B_{\mathrm{ac}}\sim\bar B_{\mathrm{ac}}$, furthermore,  $\left|\omega-\bar\omega\right|\gg
%\frac{B_\textrm{ac}^2}{\omega_\textrm{split}}
\Omega_\text{Rabi,bi}
$. The latter implies that taking the limit $\omega=\bar\omega$  in  our formulas does not give correct results for the monochromatic half-harmonic (two-photon) resonance; a similar effect is discussed in Sec.~\ref{sec:bichromaticedsr}. The reason for this limitation is the following: During the perturbative analysis of the bichromatic driving we assume, that $\omega+\bar\omega$ is matching with the qubit splitting, but $2\omega$ and $2\bar\omega$ are not. If $\omega$ and $\bar\omega$ are very close to each other, then $2\omega$ and $2\bar\omega$ are almost matching with qubit splitting, therefore we have to take into account the corresponding extra terms during the perturbative calculation.

The bichromatic Rabi frequency  is 
proportional to the product of the strength of the two driving fields. It implies that the coherent oscillation is induced only by the interplay of the two  ac fields.  The bichromatic Rabi frequency is given by a second-order formula, which is an order of magnitude smaller than the fundamental monochromatic Rabi frequency. It indicates that the bichromatic control of spin qubit  is much slower than the standard monochromatic  control, which is a drawback of this technique.
As another consequence, the power broadening of the resonance peak is smaller in case of bicromatic driving, hence more precise tuning of the driving frequencies is required to set the resonance.

As shown in Eqs. (\ref{ESR_resonance_condition}) and (\ref{ESR_BS}), the right-hand side of the resonance condition is shifted by $\omega_\text{BS}$, which is a second-order positive correction, known in the literature as the drive-strength-dependent Bloch-Siegert
shift.  It is of the same order of magnitude as the bichromatic Rabi frequency, therefore this must be considered while tuning the system into  the resonance.

\section{Bichromatic driving of the charge qubit} \label{bichromatic_charge}

In this section,
we derive and discuss the properties of the electric field-induced bichromatic resonance for a charge qubit. Note that such a setup has been investigated in the experiment in Ref. \onlinecite{PhysRevB.92.245422}, where a charge qubit was modulated simultaneously by two different ac electric fields. Subharmonic microwave resonances were already demonstrated for charge qubit defined in quantum-well structures\cite{Oosterkamp1998, VANDERWIEL199931}.
The terminology in this section refers to vertical double-dot arrangement drawn in Fig.~\ref{fig:setup}a, but the model applies equally well to the lateral arrangement of Fig.~\ref{fig:setup}b as well.

A charge qubit is defined on a single-electron double quantum dot by the following Hamiltonian:
\begin{equation} \label{eq:charge}
    H_{\textrm{charge}}=t_0\tau_1+\dfrac{\epsilon}{2}\tau_3,
\end{equation}
where  $\epsilon$ is the detuning, and $t_0$ is the tunneling amplitude. Furthermore,  $\tau_1,\;\tau_2,\;\tau_3$ are Pauli matrices acting on the basis states, that the electron occupies the top or bottom quantum dot. The energy splitting of the charge qubit is $\hbar\omega_\textrm{split}=\sqrt{4t_0^2+{\epsilon}^2}$. To describe the modulation of the gate voltage of the bottom quantum dot with frequency $\omega$ and $\bar\omega$, we augment the Hamiltonian via the modulated detuning by the following term
  \begin{equation} \label{eq:H_E2}
      H_E=\left(\dfrac{ed{F}_{\textrm{ac}}}{2}\cos{{\omega}t}+\dfrac{ed\bar{F}_{\textrm{ac}}}{2}\cos{{\bar\omega}t}\right)\tau_3,
  \end{equation}
where the distance between the quantum dots in a charge qubit is denoted by $d$, and  the amplitude of the ac electric fields are  ${F}_{\textrm{ac}}$ and $\bar{F}_{\textrm{ac}}$.  For compact notation, we introduce new variables,  $E_{\textrm{ac}}=edF_{\textrm{ac}}$ and $\bar E_{\textrm{ac}}=ed\bar F_{\textrm{ac}}$.  In contrast to the bichromatic ESR discussed in Sec.~\ref{Bichromatic_ESR}, here, the Hamiltonian  Eq. (\ref{eq:H_E2}) and consequently, the formula for the Rabi frequency [Eq. (\ref{eq:doubleRabi})] and for the Bloch-Siegert shift [Eq.(\ref{eq:doubleBloch})] are invariant under the  exchange of the two ac fields, i.e. ${E}_{\textrm{ac}},\omega \leftrightarrow \bar{E}_{\textrm{ac}},\bar{\omega}$.

With a unitary transformation, this model can be mapped to a bichromatic ESR model, where the ac fields have the same polarization. For a general value of the detuning $\epsilon$, the ac fields have both longitudinal and orthogonal components that lead to Rabi oscillation as described in the previous section. However, for $\epsilon=0$, both ac fields are orthogonal, which allows only odd number of photon processes. Therefore, the bichromatic Rabi frequency vanishes at the charge degeneracy point ($\epsilon=0$).

In case of satisfying the following conditions:
\begin{eqnarray}
\{E_\textrm{ac},\bar E_\textrm{ac}\}&\ll&\hbar\cdot\{\omega_\textrm{split},\omega,\bar\omega\},\label{weakelectric}\\
E_\textrm{ac} &\sim& \bar E_\textrm{ac},\\
\Omega_\textrm{Rabi,bi}&\ll&\left|\omega-\bar\omega\right|\label{almostsame}
\end{eqnarray}
we can use the same calculation method as in the previous section to derive the Rabi frequency 
 \begin{equation}\label{eq:doubleRabi}
  \Omega_\textrm{Rabi,bi}=\frac{E_{\textrm{ac}}\bar E_{\textrm{ac}}t_0\epsilon}{2\hbar^4\omega\bar\omega\omega_\textrm{split}},
  \end{equation}
and the Bloch-Siegert shift
\begin{equation}\label{eq:doubleBloch}
  \omega_{\textrm{BS}}=\frac{t_0^2}{\hbar^4\omega_\textrm{split}} \left(\frac{E_{\textrm{ac}}^2}{\omega_\textrm{split}^2-\omega^2}+\frac{\bar E_{\textrm{ac}}^2}{\omega_\textrm{split}^2-\bar\omega^2}\right),
\end{equation}
for bichromatic driving. The resonance condition is the same as in Eq. (\ref{ESR_resonance_condition}).

From a quantum computing perspective, it is important to perform fast Rabi oscillations as fast single-qubit gates. Hence we analyze the parameter dependence of the Rabi frequency induced by bichromatic driving. Beyond the trivial option, to increase the driving strength, in an experiment it is feasible to control the detuning and the difference of the ac frequencies. Due to the resonance condition, the sum of the two ac frequencies are fixed, but the difference is a freely variable parameter. According to Eq.~(\ref{eq:doubleRabi}) the bichromatic Rabi frequency is proportional to $\sim\frac{1}{\omega\bar\omega}$, hence this Rabi frequency has a minimum, when the ac frequencies are close to each other.  The larger the difference in ac driving frequencies, the larger the Rabi frequency of the charge qubit. However, this difference cannot be arbitrarily large, because we need to satisfy the condition of Eq. (\ref{weakelectric}). 

How to set the detuning $\epsilon$ to maximize the Rabi frequency? The bichromatic Rabi frequency is proportional to $\epsilon$, therefore in the charge degeneracy point ($\epsilon=0$) we cannot drive the charge qubit in the bichromatic way. This is in contrast with the monochromatic driving of the charge qubit, where the charge degeneracy point is the optimal detuning point that maximizes the Rabi frequency. If the ratio of the two ac frequencies is fixed, then the bichromatic Rabi frequency is proportional to $\frac{\epsilon}{\omega_\textrm{split}^3}$, which has a maximum at $\epsilon=\pm\sqrt{2}t_0$. 

Similar to the bichromatic ESR, the Bloch-Siegert shift and the Rabi frequency are given by second-order expressions. Hence, the power broadening of the resonance is comparable to the Bloch-Siegert shift, hence it should be relatively easy to measure the latter in an experiment.

\section{Bichromatic driving of the flopping-mode spin qubit} \label{bichromatic_EDSR}

In this section, as a third and last example for the bichromatic driving, we present bichromatic EDSR. As a concrete example we study the manipulation of a flopping-mode spin qubit by bichromatic electric field.  In a flopping-mode spin qubit, a spinful electron is trapped in a double quantum dot and the spin state is split by a homogeneous magnetic field.  The spin qubit can be controlled  by  ac  electric  field,  if  an  inhomogeneous magnetic  field is applied, or if the  spin-orbit  interaction  is  significant. In this section, we focus on the latter one. In case of monochromatic driving, flopping-mode electric dipole spin resonance was already studied theoretically \cite{PhysRevB.100.125430, PhysRevResearch.3.013194} and also demonstrated experimentally \cite{PhysRevResearch.2.012006}. 

\subsection{Flopping-mode spin qubit}

In a flopping-mode spin qubit, the spin and charge dynamics of the electron can be described by the following Hamiltonian
\begin{equation} \label{eq:Hamilton_flop}
  H_\textrm{flop}=t_0\tau_1+\dfrac{\epsilon}{2}\tau_3+\dfrac{1}{2}\hbar\left(\omega_z+\delta\omega_z\tau_3\right)\otimes\sigma_z + \hbar
  \tau_2 \otimes  \boldsymbol\Omega\boldsymbol\sigma,
\end{equation}
where the first two terms were already defined in Eq.~(\ref{eq:charge}). Here, we introduce $\hbar\omega_0=\sqrt{4t_0^2+\epsilon^2}$.  In our model, we take into account the $g$ factor difference between the two dots, therefore the Hamiltonian contains a dot-independent  Zeeman energy $\hbar\omega_z=g_\textrm{s}\mu_BB$  and a dot-dependent antisymmetric Zeeman energy $\hbar\delta\omega_z=g_\textrm{a}\mu_BB$, where $g_{\textrm{s/a}}=\frac{g_2\pm g_1}{2}$ are the symmetric and the anti-symmetric components of g-factors, defined from the $g$ factors $g_1$ and $g_2$ of the bottom and top quantum dots, respectively. As before, $\mu_B$ is the Bohr-magneton, $B$ is magnetic field directed in the $z$ axis, and $\sigma_x,\sigma_y$ and $\sigma_z$ are Pauli matrices in the spin subspace. $\tau$ matrices are already defined after Eq. (\ref{eq:charge}).  The last term of Eq.~\eqref{eq:Hamilton_flop} describes spin-dependent tunneling due to the spin-orbit interaction \cite{PhysRevB.80.041301, Schreiber2011, PhysRevB.102.195418}. In general, it can be characterized by three independent  parameters $  \boldsymbol\Omega=(\Omega_x,\Omega_y,\Omega_z)$. However, as we show below, in the leading-order results, only the combination $\Omega_\textrm{SO}=\sqrt{\Omega_x^2+\Omega_y^2}$ of these parameters appears.

In the following perturbative calculation, we assume this hierarchy between the parameters:
\begin{equation} \label{condition}
    \Omega_z\lesssim\Omega_\textrm{SO}\sim\delta\omega_z\ll(\omega_0-\omega_z),
\end{equation}
where the spin-orbit interaction $\Omega_\text{SO}$ and the dot-dependent Zeeman-energy $\delta \omega_z$ are taken into account as perturbations. This condition can be satisfied for certain host materials, e.g., for electrons in silicon, where $\Omega_\textrm{SO}$ is around a few $\mu$eV, and the $g$-factor antisymmetry is small, $g_a\ll g_s$\cite{PhysRevB.92.201401}. However, for other host materials with large intrinsic spin-orbit coupling strength and strong $g$-factor differences, such as germanium, the condition Eq. (\ref{condition}) is violated, and our perturbative results are inaccurate. Nevertheless, we carry out the perturbative description, because it provides a simple, intuitive physical picture, and yields analytical results that can serve as benchmarks for future numerical and experimental studies.
Note that our perturbative results apply to electronic qubits in silicon only if the energy splitting associated to the valley degree of freedom\cite{RevModPhys.85.961} exceeds the energy scales of Eq. (\ref{eq:Hamilton_flop}). 

\begin{figure}
\centering
\includegraphics[width=0.9\columnwidth]{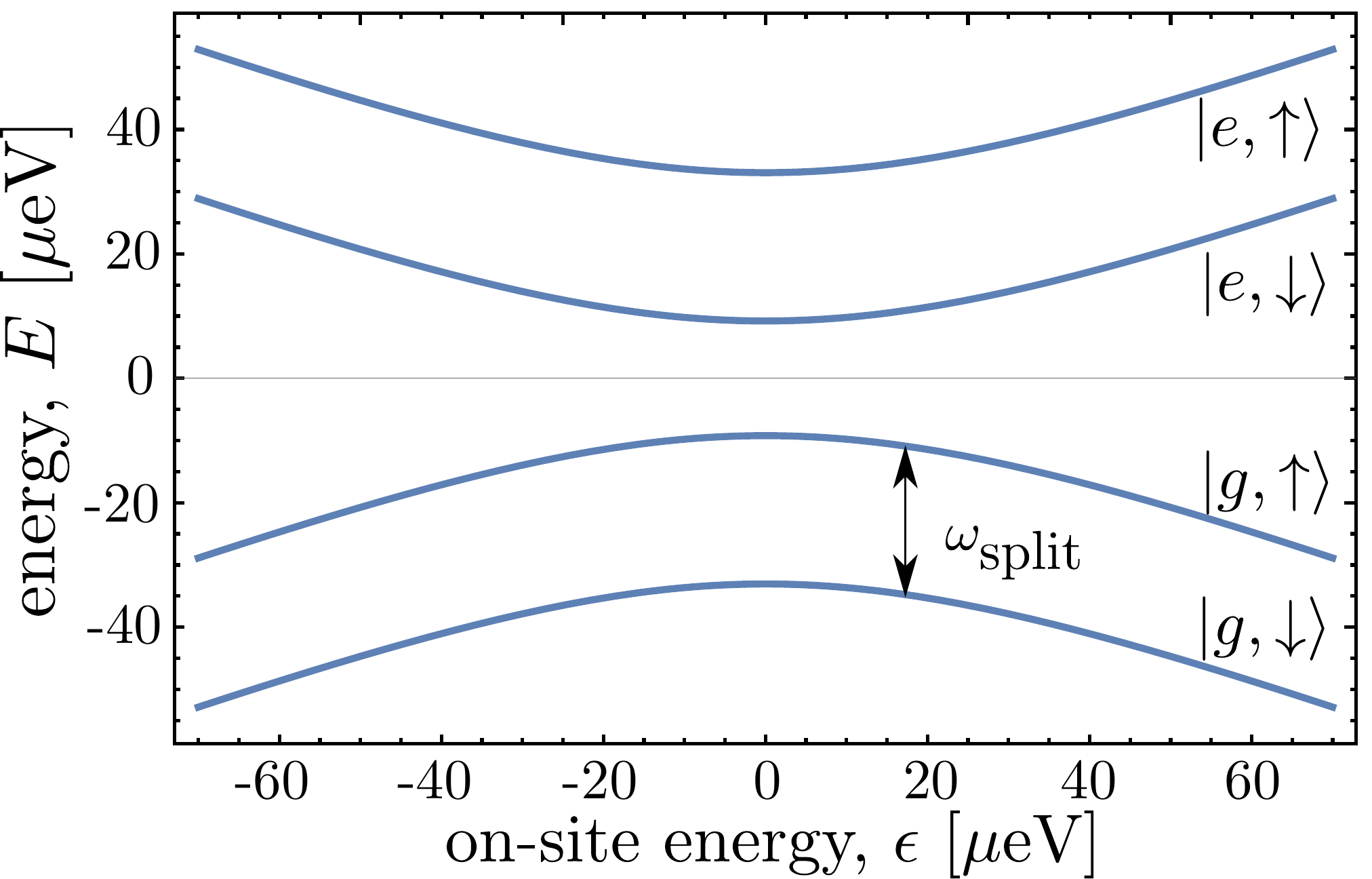}
\caption{\label{fig:flopping_level} 
Energy levels of the flopping-mode Hamiltonian as a function of the detuning. The flopping-mode spin qubit with splitting $\omega_\textrm{split}$ is formed by the two lowest energy states. On the right-hand side we show the dominant charge ($g$round or $e$xcited) and spin (up or down) configurations of the states.  Parameters:  $t_0=21\, \mu$eV, $\hbar\omega_z=24 \,\mu$eV, $\hbar\Omega_\textrm{SO}=2 \,\mu$eV, $\hbar\delta\omega_z=0 \,\mu$eV, $\hbar\Omega_z=0 \,\mu$eV.}
\end{figure}

In Fig. \ref{fig:flopping_level} the spectrum of the Hamiltonian Eq. (\ref{eq:Hamilton_flop}) is shown  as a function of detuning $\epsilon$. If the condition  Eq. (\ref{condition}) is fulfilled, then the bands, in leading order, can be labeled by the spin and the charge configurations of the states. We focus on the case $2t_0 > \hbar \omega_z$, and define the flopping-mode spin qubit basis states as the ground state and the first excited state of the Hamiltonian. The energy splitting of the flopping mode qubit $\hbar \omega_\textrm{split}$ is dominated by the dot-independent Zeeman term. However, the antisymmetric Zeeman energy and the spin-orbit coupling strength gives first-, second- and third-order corrections to that,
\begin{eqnarray} \label{split}
    \omega_\textrm{split}&\approx&\omega_z-\frac{\epsilon}{\hbar\omega_0}\delta\omega_z-\frac{2\omega_z}{\omega_0^2-\omega_z^2}\Omega_\textrm{SO}^2+\nonumber\\&+&\frac{2\delta\omega_z\epsilon(t_0^2\delta\omega_z^2+\hbar^2\omega_0^2\Omega_z^2)}{\hbar^3\omega_0^5}.
\end{eqnarray}
For a consistent description of the resonance condition, we need the third-order expansion in Eq.~\eqref{split}, since the leading-order result for the bichromatic Rabi frequency is also of third order, see next subsection.

%In the expansion of the splitting we have to go to third-order correction, because, as we will see in the next subsection, the two-photon Rabi frequencies will be derived also by a third-order perturbative calculation.

Before discussing the bichromatic Rabi oscillation, first we review the monochromatic Rabi oscillation, when the qubit is controlled by a single sinusoidal electric field, $ H_E=\dfrac{E_{\textrm{ac}}}{2}\cos\left({{\omega}t}\right)\tau_3$. If the frequency of the ac electric field matches the splitting of the flopping-mode spin qubit, i.e., $\omega=\omega_\textrm{split}$, then a fundamental monochromatic Rabi oscillation is triggered. In case of weak driving
\begin{equation} \label{weakdrive}
E_\textrm{ac}/\hbar\sim \Omega_\textrm{SO}\sim\delta\omega_z
\end{equation} the corresponding Rabi frequency can be derived by a second-order perturbative calculation
\begin{equation} \label{Rabi1}
    \Omega_{\textrm{Rabi,mono}}=\frac{2t_0\omega_z\Omega_\textrm{SO}}{\hbar^2\omega_0(\omega_0^2-\omega_z^2)}E_\textrm{ac}.
\end{equation}
This formula implies that the interplay of the electric field and the spin-orbit interaction is required for monochromatic  electrical control of the flopping-mode spin qubit. The Rabi oscillation can be sped up by tuning the system to the charge degeneracy point ($\epsilon=0$). If the tunneling amplitude $t_0$ is large compared to the Zeeman splitting, then the different  spin states are hardly hybridized, and the electric control is ineffective. Hence, for fast electrical control, the charge-qubit splitting $\omega_0$ should be tuned slightly above the Zeeman splitting $\omega_z$.

 In case of the fundamental resonance, the Bloch-Siegert shift is negligible, because it gives only a higher-order correction to the resonance condition. 

Note that our formula for the Rabi frequency Eq. (\ref{Rabi1}) at $\epsilon=0$ is almost identical to Eq. (3) of Ref. \onlinecite{PhysRevB.100.125430}. Minor difference arises from the fact that our model incorporates spin-orbit interaction, whereas the model in Ref. \onlinecite{PhysRevB.100.125430} describes an inhomogeneous magnetic field.

\subsection{Bichromatic driving}
\label{sec:bichromaticedsr}

As one of our main results, we now present our perturbative analytical expressions for the Bloch-Siegert shift and for the Rabi frequency, for the case when the flopping-mode spin qubit is controlled by bichromatic ac fields with frequency $\omega$ and $\bar\omega$ according to Eq.~(\ref{eq:H_E2}). 
We impose the condition of Eqs. (\ref{weakelectric})-(\ref{almostsame}), (\ref{condition}), and (\ref{weakdrive}). If the frequencies of the electric fields satisfy the resonance condition  $\omega+\bar\omega=\omega_\textrm{split}+\omega_\textrm{BS}$, 
then the flopping-mode spin qubit performs Rabi oscillations with the Rabi frequency
\begin{widetext}
\begin{equation}\label{double:Rabi}
  \Omega_\textrm{Rabi,bi}=\frac{t_0\epsilon\omega_z\left(\omega^2-\omega\omega_z+\omega_z^2-3\omega_0^2\right)}{\hbar^4\omega_0\left(\omega_0^2-\omega^2\right)\left(\omega_0^2-\omega_z^2\right)\left(\omega_0^2-(\omega_z-\omega)^2\right)}\Omega_\textrm{SO}E_\textrm{ac}\bar E_\textrm{ac},
\end{equation}
with a corresponding Bloch-Siegert shift
\begin{equation}\label{double:BS}
    \omega_\textrm{BS}=\frac{\epsilon t_0^2\left(3\omega_0^2-\omega^2\right)}{\hbar^5\omega_0^3\left(\omega_0^2-\omega^2\right)^2}\delta\omega_z E_\textrm{ac}^2+\frac{\epsilon t_0^2\left(3\omega_0^2-(\omega_z-\omega)^2\right)}{\hbar^5\omega_0^3\left(\omega_0^2-(\omega_z-\omega)^2\right)^2}\delta\omega_z \bar E_\textrm{ac}^2.
\end{equation}
\end{widetext}
These formulas were derived by the same technique as in the previous sections: many-mode Floquet-theory combined with Schrieffer-Wolff transformation. 

\begin{figure}
\centering
\includegraphics[width=0.95\columnwidth]{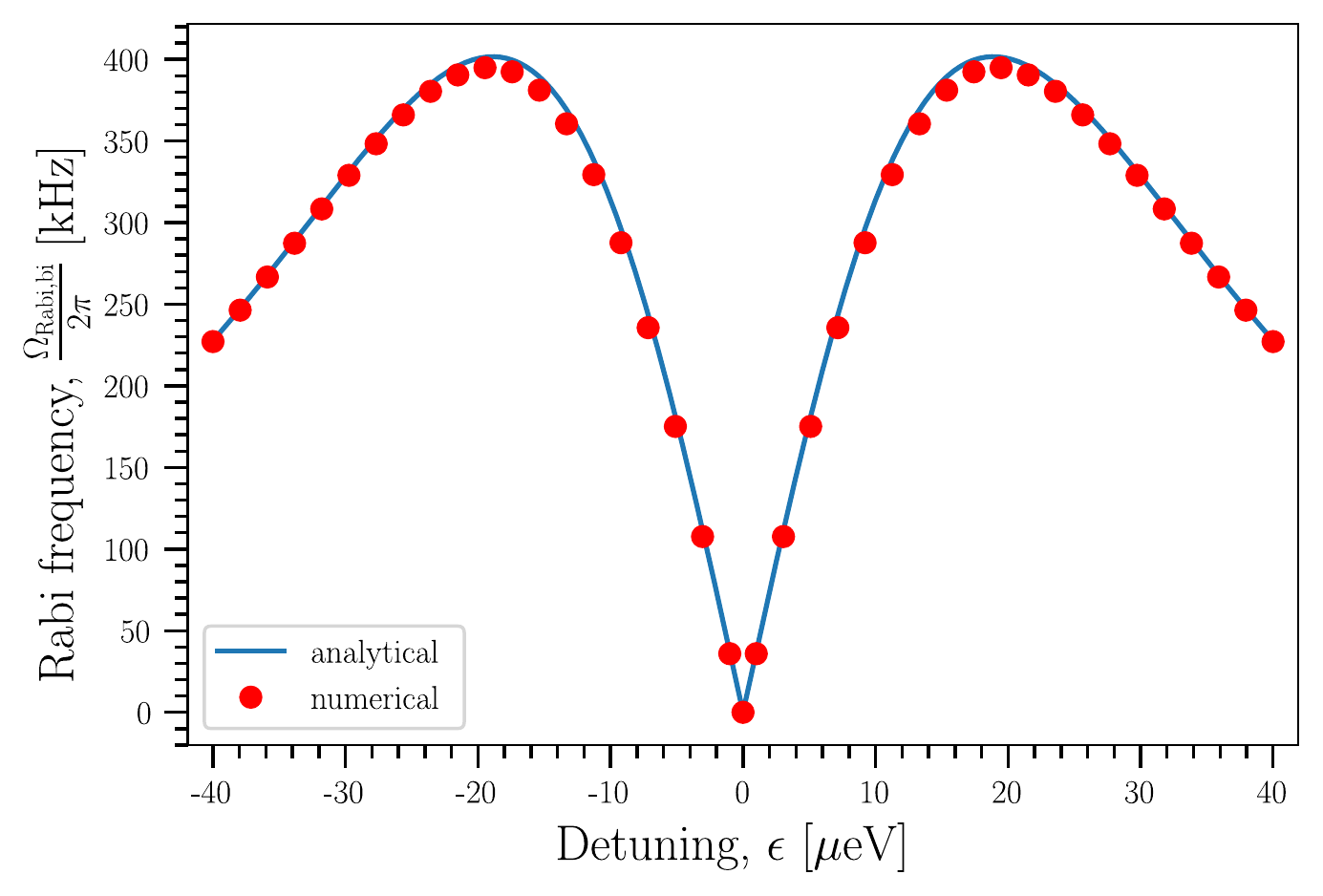}
\caption {\label{fig:flop-epsilon} Bichromatic Rabi frequency as function of detuning for the flopping-mode spin qubit. Blue-solid  line: Analytical result (\ref{double:Rabi}). Red dots: Result from the numerical simulation, where $\bar\omega$ was optimised numerically for each $\epsilon$ value to obtain complete Rabi oscillations.
Parameters: $t_0=21\,\mu$eV, $\hbar\omega_z=24 \,\mu$eV, $\hbar\Omega_y=2 \,\mu$eV, $\hbar\Omega_x=\hbar\Omega_z=0\, \mu$eV, $\hbar\delta\omega_z=0\,\mu$eV, $\omega=0.7 \omega_\textrm{split}$, and $E_\textrm{ac}=\bar E_\textrm{ac}=2 \,\mu$eV.}
\end{figure}

\begin{figure}
\centering
\includegraphics[width=0.95\columnwidth]{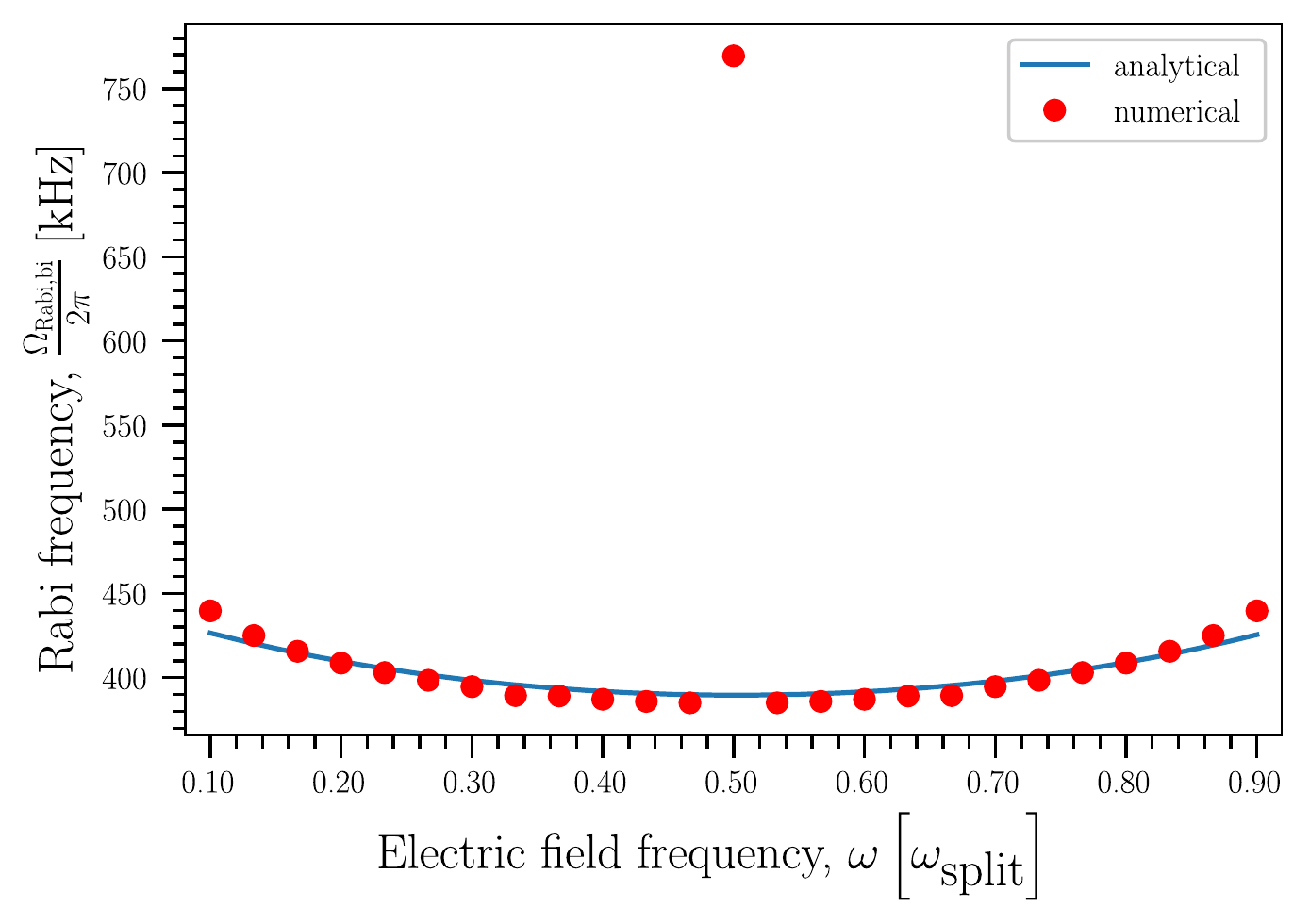}
\caption {\label{fig:wwbar_fugges}  Bichromatic Rabi frequency  as function of the frequency of  one  of  the  driving  fields, $\omega$.   Blue-solid  line: Analytical result (\ref{double:Rabi}). Red dots: Result from the numerical simulation, where  $\bar\omega$ was optimised numerically for each $\omega$ value to obtain complete Rabi oscillations.
Parameters: $\epsilon=21 \,\mu$eV, $t_0=21\,\mu$eV, $\hbar\omega_z=24\,\mu$eV, $\hbar\Omega_\textrm{SO}=2 \,\mu$eV, $\Omega_z=0 \,\mu$eV, $\hbar\delta\omega_z=0 \,\mu$eV, and $E_\textrm{ac}=\bar E_\textrm{ac}=2 \,\mu$eV.}
\end{figure}

\begin{figure}
\centering
\includegraphics[width=0.95\columnwidth]{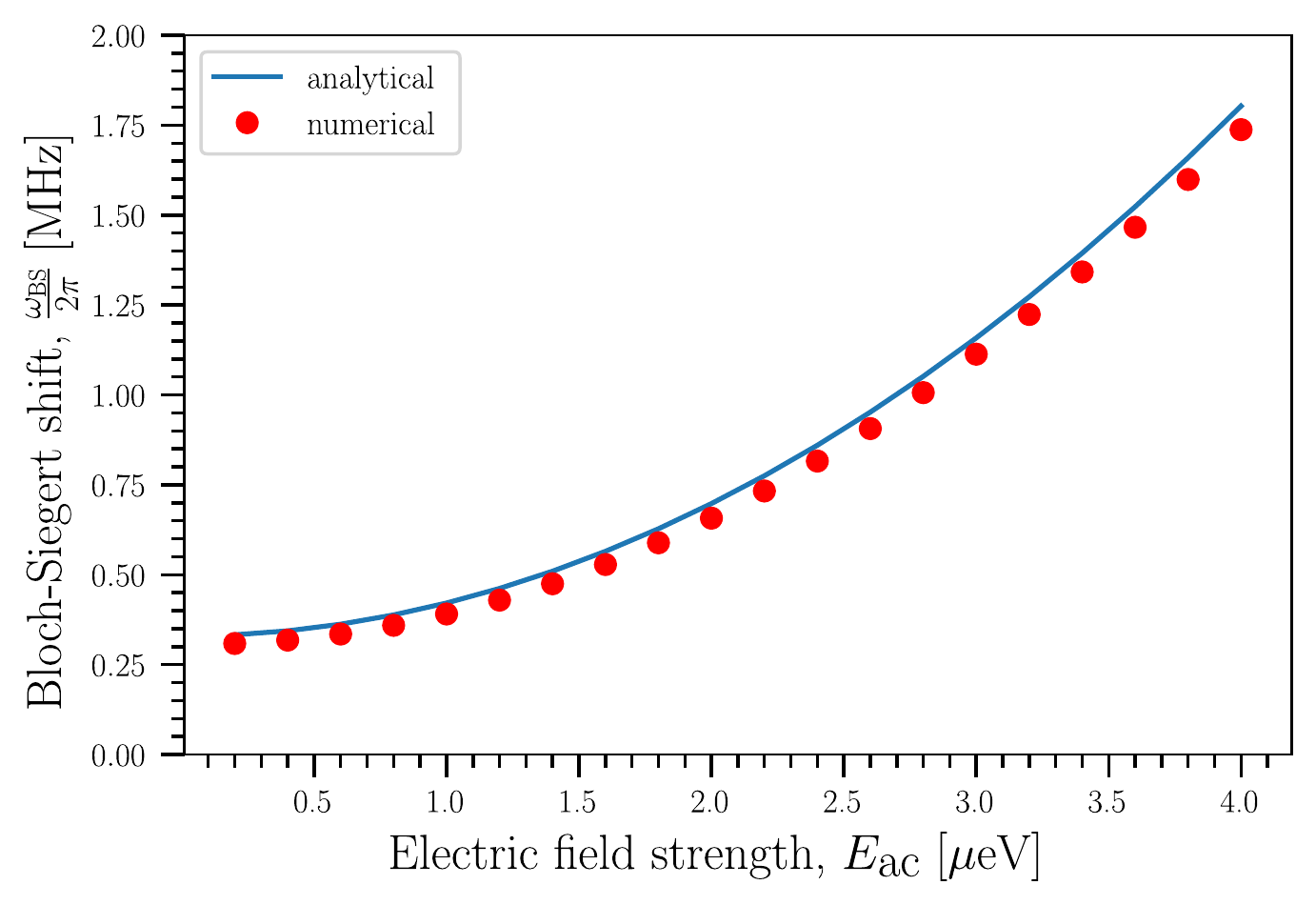}
\caption {\label{fig:wBS} Bichromatic Bloch-Siegert shift as function of the strength of one of the two electric fields $E_\mathrm{ac}$. Blue-solid line: Analytical result (\ref{double:BS}). The other electric field strength is fixed, $\bar E_\mathrm{ac}=2 \, \mu$eV. The remaining parameters: $\hbar\omega_z=24\,\mu$eV, $\hbar\Omega_y=2\,\mu$eV, $\hbar\Omega_x=\hbar\Omega_z=0\,\mu$eV, $\epsilon=30\,\mu$eV, $t_0=21\,\mu$eV, $\hbar\delta\omega_z=3\,\mu$eV, and $\omega=0.7\omega_\mathrm{split}$. $\omega_\mathrm{split}$ was determined numerically, $\omega_\mathrm{split}/(2\pi)=5.36026$ GHz. $\bar\omega$ was optimised numerically for each $E_\textrm{ac}$ value to obtain complete Rabi oscillations}
\end{figure}

Let us discuss the relevant features of our key analytical results Eqs.~\eqref{double:Rabi} and \eqref{double:BS}. The proportionality $\Omega_\textrm{Rabi,bi}\sim\Omega_\textrm{SO}E_\textrm{ac}\bar E_\textrm{ac}$ signals that a third-order perturbation theory was used during the derivation, and the Rabi oscillation is the result of the interplay of the spin-orbit interaction and the electric fields. The $x$ and $y$ components of the spin-orbit interaction  are key ingredients of the coherent spin rotation, because besides that no other spin flopping term is built in our model. 

On-site energy detuning $\epsilon$ is one of the tunable parameters in an experiment, hence we plot the  Rabi frequency as a function of the detuning  in Fig.~\ref{fig:flop-epsilon}. Here we support our analytical result (solid blue) with numerical results (red points) obtained via solving the time-dependent Schrödinger-equation with the Runge-Kutta method. According to  Eq.~(\ref{double:Rabi}), the Rabi frequency is zero in the charge degeneracy point, and it has a maximum at a finite energy detuning, where $\epsilon$ is of the order of $t_0$ and $\omega_z$. For the exact position of the maximum we need to maximize the analytical formula \eqref{double:Rabi} with respect to $\epsilon$.

From an experimental viewpoint, setting the working point at the vicinity of the maximum has two advantages: first, it ensures a high Rabi frequency; second, the Rabi frequency is insensitive to charge noise at this point in first order, which fosters long-lived Rabi oscillations in the presence of charge noise. 

%of view  it is worth to control the system close to the maximum, because we can increase the speed of the oscillation, furthermore,  we can minimize the harmful effect of the charge noise, because here in linear order the Rabi frequency is insensitive to the fluctuation of the detuning. 

In case of bichromatic driving, the resonance condition fixes the sum of $\omega$ and $\bar\omega$, but there is a freedom to divide the sum between $\omega$ and $\bar\omega$. In Fig. \ref{fig:wwbar_fugges}, the  Rabi frequency is plotted as a function of this division, namely, as a function of $\omega$. Here, the analytical results are also tested against numerically
exact solutions of the time-dependent Schr\"odinger equation.  The figure shows that for an experimentally motivated parameter set, the Rabi frequency is hardly controllable by changing the division; only a slight increase of the Rabi frequency is observed as the difference $|\omega-\bar\omega|$ is increased.  

The numerical results in Fig. \ref{fig:wwbar_fugges} reveal the limitations of our analytical method. 
In the center of the graph, the stand-alone numerical data point illustrates the breakdown of perturbation theory in the range $\omega \approx \bar{\omega}$, where the condition Eq. (\ref{almostsame}) is violated.
Note that the numerically obtained Rabi frequency is approximately twice as large as the analytical result indicates; the numerical result can be recovered by an appropriate perturbative description of the half-harmonic resonance\cite{PhysRevB.92.054422}.
At the edges of the plot, where the condition Eq. (\ref{weakelectric}) is violated, the discrepancy between the analytical and numerical result increases. For even more extreme values of $\omega$, the Rabi oscillations become distorted (not shown).

%respects to not a bichromatic but a monochromatic as well as first-subharmonic driving, when $\omega=\bar\omega\approx\frac{\omega_\textrm{split}}{2}$, so the condition Eq. (\ref{almostsame}) is violated. Therefore, the corresponding frequency cannot be described by Eq.~(\ref{double:Rabi}), the former is bigger by a factor of two. 

The proportionality $\omega_\textrm{BS}\sim\delta\omega_z E_\textrm{ac}^2$ in Eq.~\eqref{double:BS} signals  that the leading order term of the Bloch-Sigert shift is  derived by third-order perturbation theory, similarly to the formula \eqref{double:Rabi} of the Rabi frequency. These two quantities are of the same order in the small parameters, i.e., the drive-strength-dependent shift of the resonant condition is as significant as the power broadening. According to Eq. (\ref{double:BS}) the Bloch-Sigert shift is due to the interplay of the $g$ factor antisymmetry and the electric field. If the $g$-tensors are symmetric in the two dots, i.e., $g_a=0$, then the Bloch-Siegert shift will be a fourth-order effect and can be neglected. In Fig.~\ref{fig:wBS}, the analytical result of the Bloch-Siegert shift Eq. (\ref{double:BS}) is verified by numerical simulation.

Let us close this subsection by discussing the role of phase shifts in the ac electric fields. So far, we investigated the scenario when the two ac fields are switched on in-phase, i.e.,  both fields have the maximal value at $t=0$. Even if we take into account the phases of the driving fields, described by the Hamiltonian
  \begin{equation} \label{eq:H_E_phase}
      H_E=\frac{1}{2}\left({E}_{\textrm{ac}}\cos{({\omega}t+\phi)}+{\bar E}_{\textrm{ac}}\cos{({\bar\omega}t+\bar\phi)}\right)\tau_3,
  \end{equation}
the resulting Rabi frequencies are insensitive to the phase shifts $\phi$ and $\bar{\phi}$. However, different phase shifts result in different  single-qubit gates. (In the description of bichromatic ESR in Appendix \ref{app_Schrieffer_Wolff} we take into account these phase shifts.)

\subsection{Charge noise}

As we have shown above, Rabi oscillations due to bichromatic driving are slower than those due to monochromatic driving, if the drive strengths are the same. In this subsection, we show the consequence that bichromatic EDSR is rather sensitive to charge noise. In particular, we illustrate the reduction of the visibility of bichromatically induced Rabi oscillations due to charge noise. 
%the quality of bichromatically induced Rabi oscillations ow, this implies, as shown below, that in the bichromatic case, the qubit is exposed mo to the harmful effect of the environmental noise.  In this subsection we demonstrate, that electrical potential fluctuation can hinder the experimental observability of bichromatic oscillation. 

\begin{figure}
\centering
\includegraphics[width=0.95\columnwidth]{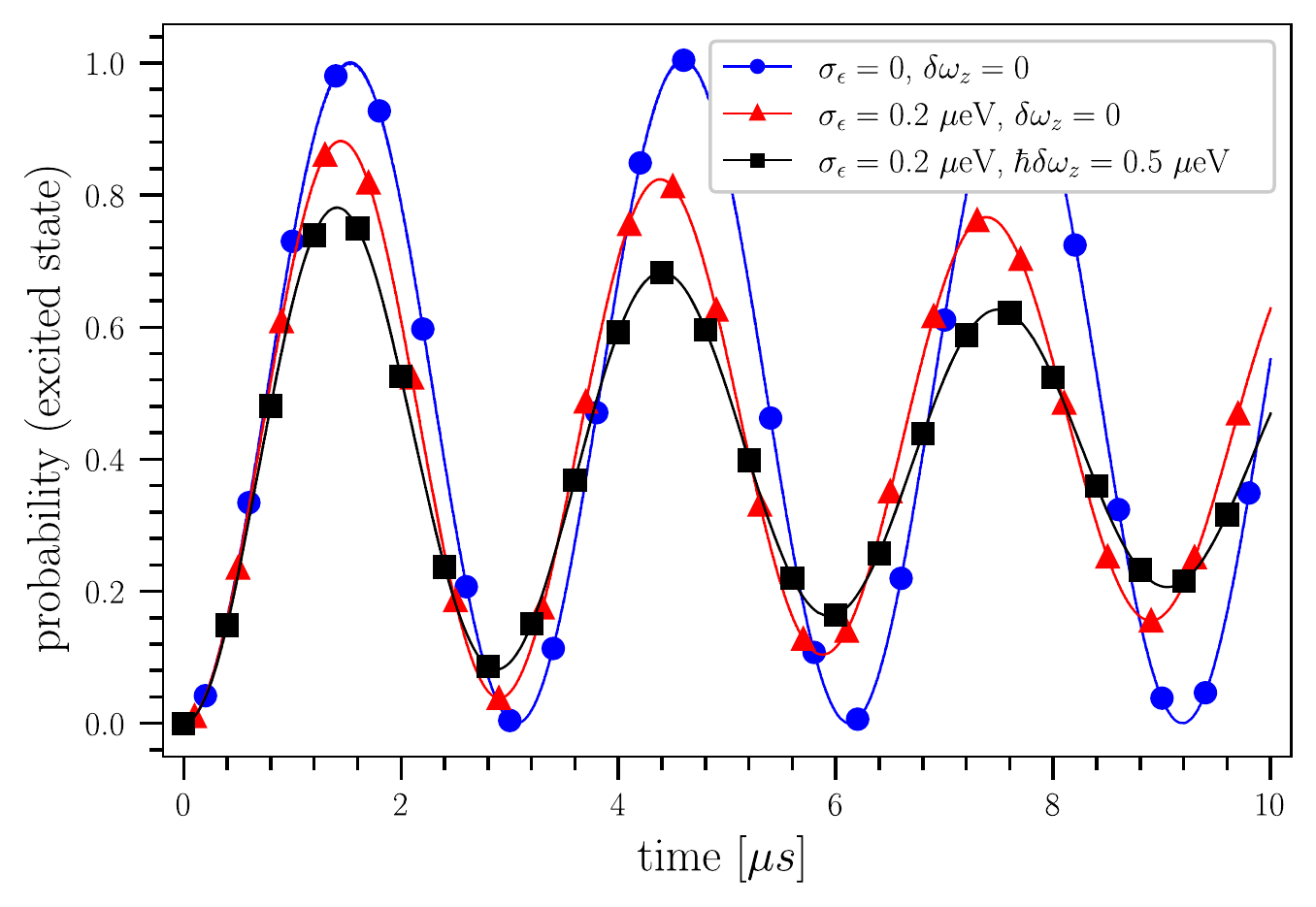}
\caption {\label{fig:charge_noise} 
Damping of the Rabi oscillation due to charge noise and $g$-factor asymmetry.  Frequency $\bar\omega$ was optimised numerically to maximize the fidelity, yielding $\bar\omega/(2\pi)=1.73442$ GHz for $\delta\omega_z=0$ (red curve), and $\bar\omega/(2\pi)=1.71347$ GHz for $\hbar\delta\omega_z=0.5\,\mu$eV (black curve). The remaining parameters: $t_0=21\,\mu$eV, $\epsilon_0=30 \,\mu$eV, $\hbar\omega_z=24 \,\mu$eV, $\hbar\Omega_y=2 \,\mu$eV, $\hbar\Omega_x=\hbar\Omega_z=0\, \mu$eV, $E_\textrm{ac}=\bar E_\textrm{ac}=2 \,\mu$eV, and $\omega=0.7\,\omega_\textrm{split}$. Splitting frequency $\omega_\textrm{split}$ was calculated from Eq. (\ref{split}).
Marks on lines are only guides to the eye.}
\end{figure}

Electrical potential fluctuation is one mechanism responsible for the damping of Rabi oscillations. Here, we model this type of noise as a random quasistatic fluctuation of the on-site detuning. Quasistatic means that the value of the detuning is assumed to be fixed for each measurement run, but changes randomly between the subsequent measurement runs. This type of noise leads to imperfect Rabi oscillations due to two reasons: (i) The Rabi frequency $\Omega_\textrm{Rabi}(\epsilon)$ itself depends on the detuning. (ii) The fluctuation detunes the system from the resonance condition, hence the qubit is driven off-resonantly.  If the ac electric field frequencies $\omega$ and $\bar{\omega}$ are calibrated resonantly before the measurement at the value of detuning $\epsilon_0$, then for an other $\epsilon$ realization the driving will be off-resonant with a frequency detuning of 
$\omega_\textrm{off}(\epsilon)=\omega_\textrm{split}(\epsilon)+\omega_\text{BS}(\epsilon)-\omega_\textrm{split}(\epsilon_0)-\omega_\text{BS}(\epsilon_0)$. 

For a certain $\epsilon$ realization, the probability that the population of the system is in its excited state is given by 
\begin{equation} \label{Pt}
    P(t,\epsilon)=\frac{\Omega_\textrm{Rabi}^2(\epsilon)}{\Omega_\textrm{Rabi}^2(\epsilon)+\omega_\textrm{off}^2(\epsilon)}\sin{\left[\frac{1}{2} t \sqrt{\Omega_\textrm{Rabi}^2(\epsilon)+\omega_\textrm{off}^2(\epsilon)} \right]^2}.
\end{equation}
For simplicity, the probability distribution of $\epsilon$ is assumed as a Gaussian distribution with mean value $\epsilon_0$ and standard deviation $\sigma_\epsilon$. The averaged excited state occupation probability is derived by averaging over the on-site energy in Eq.~(\ref{Pt}), yielding
\begin{equation} \label{int}
    \overline{P}(t)=\int\displaylimits_{-\infty}^{\infty}d\epsilon P(t,\epsilon) \frac{1}{\sqrt{2\pi}\sigma_\epsilon}e^{-\frac{(\epsilon-\epsilon_0)^2}{2\sigma_\epsilon^2}}.
\end{equation}
The damping rate of  $\overline{P}(t)$  is characterized by its maximal value $\mathcal{F}=\max_t \overline{P}(t)$, which we call the \emph{fidelity}. In fact, this quantity describes the fidelity of a $\pi$ rotation around the $x$ axis (i.e., an $X$ gate), if this gate is acting on the ground state.

The damped Rabi oscillation $\overline{P}(t)$ is plotted in Fig.~\ref{fig:charge_noise} for different strength of charge noise and $g$-factor asymmetry. The $\bar P(t)$ was calculated using Eq. (\ref{int}), where $P(t,\epsilon)$ was evaluated numerically, solving the Schr\"odinger-equation for 100 different $\epsilon$ values evenly distributed in $[\epsilon_0-4\sigma_\epsilon,\epsilon_0+4\sigma_\epsilon]$ interval, approximating the integral with a sum. 

A key observation is that a relatively small charge noise $\sigma_\epsilon=0.2$ $\mu$eV \cite{https://doi.org/10.48550/arxiv.2104.03045}, even in a symmetric double quantum dot $g_a=0$, decreases the fidelity significantly, to approximately 0.9 (red triangles in Fig.~\ref{fig:charge_noise}).  Furthermore, the frequency detuning $\omega_\textrm{off}(\epsilon)$ depends strongly on the antisymmetric Zeemann term $\delta \omega_z$, because the leading order $\epsilon$-dependent term of both the Bloch-Siegert shift and the qubit splitting is proportional to $\delta\omega_z$. Therefore, the damping rate increases significantly, when the $g$ factors are different in the quantum dots. This is shown  by the black curve in Fig.~\ref{fig:charge_noise}, where small $g$-factor difference leads to a fidelity less than 0.8. Observation of bichromatic Rabi oscillation requires reduced charge noise and a reduced asymmetry of the $g$ factors. 

Charge-noise resilience  can be improved by tuning the system into dynamical sweet spots, where, up to leading order, both the Rabi frequency and the frequency detuning $\omega_\textrm{off}(\epsilon)$ are insensitive to the charge noise.
We should note that the idle sweet spot, where the system is protected against dephasing, differs from the dynamical one. The former one is at the charge degeneracy point $\epsilon = 0$, where we cannot drive the system, because the Rabi frequency is zero. For monochromatic driving, a careful analysis of the sweet spots were done in Ref. \onlinecite{PhysRevB.100.125430}, while for bichromatic driving this could be a subject of future work.

\section{Applications} \label{bichromatic_application}

Shared control of a 2D spin qubit array, as envisioned in Ref.~\onlinecite{doi:10.1126/sciadv.aar3960}, provides a strong, square-root reduction ($N_c \propto \sqrt{N_q}$) of the number of control lines $N_c$ with respect to the qubit count $N_q$. Furthermore, such a setup could serve as a platform for fault-tolerant quantum computing \cite{Helsen_2018}.
In this section, we highlight the advantages of electric field-induced bichromatic driving in a 2D grid of semiconducting quantum dots with shared control.

Instead of building up separate control lines to each  quantum dot, shared control is envisioned by a crossbar architecture \cite{doi:10.1126/sciadv.1500707,doi:10.1126/sciadv.aar3960}. In such a setup, on-site energies of the quantum dots can be controlled by a parallel set of long plunger gates, in such a way that one plunger gate controls a row or a column of quantum dots. In the vertical and lateral arrangements, sketched in  Fig. \ref{fig:setup}, there are two sets of long and straight parallel plunger gates, the two sets being perpendicular to each other. As discussed below, bichromatic driving enables selective addressing and parallelized single-qubit gates in such 2D qubit arrays.

In the vertical arrangement shown in Fig.~\ref{fig:setup}(a), the double dots are perpendicular to the plane of the 2D array. This may be created by a 3D-fabrication technique\cite{doi:10.1126/sciadv.1500707}. An alternative setup, inspired by Ref.~\onlinecite{doi:10.1126/sciadv.aar3960}, is shown in Fig. \ref{fig:setup}(b). There, the 2D-quantum dot array is defined by the barrier gates shown as horizontal and vertical lines in Fig.~\ref{fig:setup}(b). Upon performing flopping-mode single-qubit gates, the grey barrier gates allow hybridization of the neighboring quantum dots, while the black ones do not. As a result, we get a grid of separated double dots. Each double dot is occupied by an electron, which hybridizes between the quantum dots, as it is shown with grey disks in Fig. \ref{fig:setup}(b) for one of double dots. This platform can be operated as an array of charge qubits or flopping-mode spin qubits. In the former case, the qubit is defined by the location of the electron, while in the latter case the qubit is defined by the spin of the electron. Detuning in the double quantum dot can be controlled by the dc voltage at the diagonal plunger gates, while single-qubit gates can be realized by ac voltage at those plunger gates.

\subsection{Selective addressing}

In the shared-control qubit arrays shown in Fig.~\ref{fig:setup}, the selective addressing of the qubits can be achieved using  bichromatic driving, when ac voltages with different frequencies ($\omega$ and $\bar{\omega}$) are applied on two perpendicular control lines. If the sum of the two frequencies matches the qubit splitting, then only the qubit at the intersection point of the ac-driven control lines is controlled. The nature of the single-qubit gate is set by the duration and phase of the ac pulses. The corresponding Rabi frequencies were calculated for a charge qubit in Sec.  \ref{bichromatic_charge}, and for a flopping-mode spin qubit in Sec.~  \ref{bichromatic_EDSR}.

The advantage of the selective addressing with bichromatic driving is that this method works reliably even if the qubits are uniform. For example, there is no need for local control of the Zeeman splitting, which would be required when addressing a single qubit with a monochromatic field. A potential disadvantage of bichromatic driving, as compared to monochromatic driving with similar strength, is the relative slowness of single-qubit gate.

\subsection {Parallelization}

\begin{figure}[t]
\centering
\includegraphics[width=0.85\columnwidth]{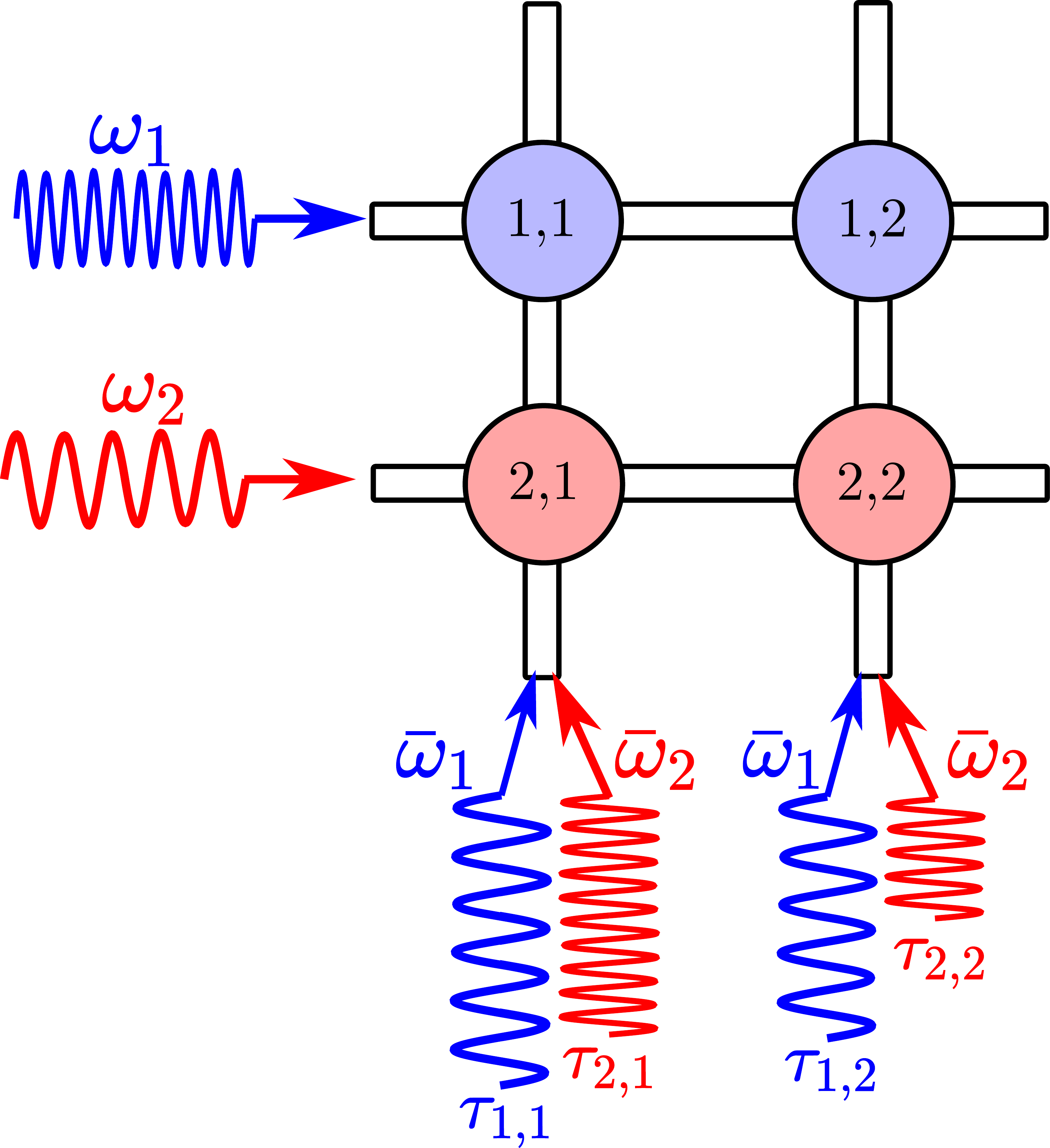}

\caption {\label{fig:para} Parallelization of single-qubit gates in a 2D-qubit array with shared control. Blue (red) qubits in the first (second) row are driven bichromatically by the blue (red) ac fields. By an appropriate choice of the durations ($\tau_{1,1}, \dots, \tau_{2,2}$)  and the initial phases of pulses through the  vertical control lines, any combination of single-qubit gate is achievable. The voltage on the vertical-control lines is the sum of ac voltage pulses with different frequencies.}
\end{figure}

Combining bichromatic driving with frequency multiplexing opens the opportunity of parallelization, i.e., the simultaneous realization of different single-qubit gates on each qubit of the 2D array. This is illustrated in Fig.~\ref{fig:para}, where qubits (disks) are identified by their row and column indices. Qubits in the first row are driven bichromatically by frequencies $\omega_1$ and $\bar\omega_1$ satisfying the resonance condition $\omega_1+\bar\omega_1=\omega_\textrm{split}$. In our example, different single-qubit gates can be performed simultaneously on each qubit of this first row by appropriately choosing the duration and initial phase of the ac pulses on the vertical control lines. We can control qubits in the second row with the same technique, but we need other driving fields ($\omega_2$ and $\bar\omega_2$)  to avoid harmful crosstalk. 

In general, in an $N\times N$ qubit grid, $2N$ different frequencies $\omega_1, \omega_2, \dots \omega_N$ and $\bar\omega_1, \bar\omega_2, \dots \bar\omega_N$ are required for a complete parallelization, such that these frequencies fulfill the resonance condition in pairs, $\omega_i+\bar\omega_i=\omega_\textrm{split}$. This can be realized such that a monochromatic ac field is applied on the horizontal control lines, and the sum of $N$ pulses with different frequencies is applied on the vertical control lines. Setting the pulse durations  ($\tau_{1,1}, \dots, \tau_{2,2}$ in Fig. \ref{fig:para}) and the initial phases of the pulses appropriately, any combination of simultaneous single-qubit gates is achievable. 

To avoid harmful crosstalk, we need to ensure that the interplay of the electric fields with frequencies $\omega_i$ and $\bar\omega_j$ ($i\neq j$) do not drive the qubits. Because of the relatively small power broadening (low Rabi frequency) of the bichromatic resonance, it is relatively easy to find such a set of drive frequencies for small $N$. However, this task becomes more challenging and frequency crowding becomes an issue as the setup is scaled up by increasing $N$.

We numerically demonstrate the parallelization in a $3\times3$ qubit grid, where the flopping-mode spin qubits are uniform, with parameters $\epsilon=30\, \mu$eV, $t_0=21\,\mu$eV, $\hbar\omega_z=24\,\mu$eV, $\hbar\Omega_\textrm{SO}=2\,\mu$eV, $\hbar\Omega_z=0 \, \mu$eV, and $\hbar\delta\omega_z=0 \,\mu$eV. Starting each qubit from the ground state, in our simulations we find each qubit flipped with at least 99.5\% fidelity, 
%that for each qubit, the effect of an $X$ gate we  are able to achieve the intended X-gate for every qubit at least with 99.5\% fidelity 
if the parameters of the pulses are chosen in the following way: $E_\textrm{ac}=2\, \mu$eV for every pulse; $\tau_{1,i}=1543$ ns, $\tau_{2,i}=1564$ ns, $\tau_{3,i}=1615$ ns ($i\in\{1,2,3\}$); $\omega_1/(2\pi)=4.62483$ GHz, $\bar\omega_1/(2\pi)=1.15620$ GHz, $\omega_2/(2\pi)=4.04677$ GHz, $\bar\omega_2/(2\pi)=1.73430$ GHz,  $\omega_3/(2\pi)=3.46874$ GHz, and $\bar\omega_3/(2\pi)=2.31230$ GHz.

Here, we describe how the numerical parameters above, yielding the protocol for simultaneous high-fidelity quantum gates, were constructed. A similar line of thought can be followed when designing qubit experiments with bichromatic shared control. First, we choose half of the drive frequencies according to $\bar{\omega}_1 = 0.2\, \omega_\text{split}$,
$\bar{\omega}_2 = 0.3\, \omega_\text{split}$,
$\bar{\omega}_3 = 0.4\, \omega_\text{split}$.
These choices imply the initial values  for the frequencies $\omega_1$, 
$\omega_2$ and $\omega_3$
(e.g., $\omega_1 = 0.8\, \omega_\text{split}$), and the 
pulse durations $\tau_{i,j}$.
The initial values of the pulse durations are estimated from the bichromatic Rabi frequency, see  Eq.~\eqref{double:Rabi}. Starting from these initial values of $\omega_i$ and $\tau_{i,j}$ we fine-tune these values to optimize the spin-flip probability, which we compute by numerically solving the time-dependent Schr\"odinger equation.

The above-described way of parallelization requires  high-precision control of frequencies and pulse durations.
If the frequencies are detuned from the optimal point (see above) by a few tens of kHz, then the fidelity drops by a few percent. For example, if the driving frequencies are given with less digits, $\omega_1/(2\pi)=4.6248$ GHz, $\omega_2/(2\pi)=4.0468$ GHz, and $\omega_3/(2\pi)=3.4687$ GHz, the fidelity drops to 97.3-97.7\%.  The Rabi-oscillation is accompanied by the so-called Bloch-Siegert oscillation, i.e.,  a fast and small amplitude oscillation is added to the   sine curve of the probability vs. time function. In case of our high fidelity optimization we took into account the Bloch-Siegert oscillation, therefore, the pulse durations are given by nanosecond precision.

\section{Conclusions}
\label{discussion}

In conclusion, we have analysed bichromatic EDSR, that is, electrically driven spin resonance with bichromatic driving. 
Our paper focuses on a single electron in a double quantum dot with spin-orbit interaction, operated as a flopping-mode spin qubit. 
We have found that the Rabi frequency is maximized, and hence the single-qubit gate times are minimized, at a nonzero detuning from the charge-qubit tipping point. 
We have also found that a $g$-factor difference between the dots (or more generally, the inhomogeneity of the effective magnetic field felt by the electron) causes two significant effects: (i) It induces a significant Bloch-Siegert shift of the resonance frequency, which is comparable to the power broadening, and (ii) It enhances the adversary effect of charge noise. We have also highlighted that bichromatic EDSR, combined with advanced frequency multiplexing techniques, enables simultaneous single-qubit gates in a crossbar-based shared-control spin qubit architecture.
We envision that our results will foster the design and interpretation of future experiments on multi-qubit registers with shared control.

\begin{acknowledgments}
We acknowledge helpful discussions and correspondence with M. Veldhorst and S. Zihlmann.
This research was supported by the Ministry of Innovation and Technology (MIT) and the National
Research, Development and Innovation Office (NKFIH)
within the Quantum Information National Laboratory
of Hungary and by
the NKFIH through the OTKA Grants FK 124723, FK 132146 and
FK 134437, and the European Union through the Horizon Europe grant IGNITE.
\end{acknowledgments}

\appendix

\section{Many-mode Floquet theory}
\label{app_multi_floquet_theory}

In order to calculate the Rabi frequencies of spin rotations and the corresponding drive-dependent shift of the resonance, the so-called Bloch-Siegert shift, we have to solve the initial-value problem suggested by the Hamiltonian defined in Eq.~(\ref{ESR_Hamiltonian}). Because of the bichromatic driving we are not able to solve this using the simple rotating wave approximation, instead we use the so-called Floquet theory, which was applied for solving the Schr\"odinger equation with monochromatic driving first by Shirley in Ref. \onlinecite{PhysRev.138.B979}. The Hamiltonian describing the system is required to be periodic in time, but it was shown that the calculation can be extended to polychromatic driving \cite{ho1983semiclassical}. Here we briefly review this extension of the Floquet theory for bichromatic driving. The time-dependent Schr\"odinger equation in matrix form is ($\hbar=1$)
\begin{equation}\label{schrodinger_equation}
    i\dfrac{\mathrm{d \Psi}}{\mathrm{d} t}=H\Psi, 
\end{equation}
where H is the Hamiltonian, 
\begin{equation}\label{general_Hamiltonian}
    H=H_0+V\cos{(\omega t+\phi)}+\bar V\cos{(\bar\omega t+\bar\phi)} . 
\end{equation}

$H_0$ is the Hamiltonian of the unperturbed, generally $d$-level system, with eigenstates $\ket{\alpha}$ and eigenvalues $E_\alpha$, where $\alpha$ $\in$ $\{0,1,\ldots , d-1\}$, $V$ and $\bar V$ are the operators that describe the driving of the system. The problem is that this Hamiltonian is not always periodic in time. This can be solved if we introduce a frequency $\delta\omega$, such that 
\begin{equation}
    \omega=N\delta\omega, \hspace{3mm} \bar\omega= \bar N\delta\omega, \hspace{3mm} 
    N,\bar N \in\mathbb Z^+ .  
\end{equation}
We can choose this $\delta\omega$ to be arbitrarily small, so that integers $N$ and $\bar N$ can be found and will give the $\omega$ and $\bar\omega$ frequencies with any desired precision. With the introduction of this new frequency $\delta\omega$ the Hamiltonian becomes periodic with period $T=2\pi/\delta\omega$ and we can apply the Floquet method. The solution of Eq.~(\ref{schrodinger_equation}) can be written in the form: 
\begin{equation}
    \Psi(t)=\Phi(t)\mathrm{e}^{-iQt},
\end{equation}
where $Q$ is represented by a constant, diagonal matrix, with diagonal elements $q_\alpha$,  furtermore, $\Phi(t)$ is represented by a periodic matrix with period $T$. If we expand the $\Phi(t)$ in Fourier series we get: 
\begin{equation}
    \Psi_{\alpha\beta}(t)= \sum_{n=-\infty}^{\infty} \Psi_{\alpha\beta}^{(n)} \mathrm{e}^{in\delta\omega t}\mathrm{e}^{-iq_\beta t},  
\end{equation}
where $\Psi_{\alpha\beta}^{(n)}$ denotes the Fourier component, the greek letter indices refer to the matrix elements of the solution $\Psi$. The Hamiltonian is also expanded in Fourier series
\begin{equation}
    H_{\alpha\beta}=\sum_{n=-\infty}^{\infty} H_{\alpha\beta}^{(n)}\mathrm{e}^{in\delta\omega t}.
\end{equation}
With the substitution of the Fourier expansions in the Schr\"odinger equation we get recursion relations for the $\Psi_{\alpha\beta}^{(n)}$. These relations are equivalent with an eigenvalue equation
\begin{equation}
    \sum_{\gamma=0}^{d-1} \sum_{k=-\infty}^{\infty} \left(H_{\alpha\gamma}^{(n-k)}+n\delta\omega \delta_{\alpha\gamma}\delta_{kn}\right)\Psi_{\gamma\beta}^{(k)}=q_\beta\Psi_{\alpha\beta}^{(n)} . 
\end{equation}
This is the eigenvalue equation of the so-called Floquet Hamiltonian $H_F$, which is now time-independent but in return infinite-dimensional. The Floquet Hamiltonian is defined the following way: 
\begin{equation}
    \bra{\alpha n}H_F\ket{\beta m}=H_{\alpha\beta}^{(n-m)}+n\delta\omega\delta_{\alpha\beta}\delta_{nm}. 
\end{equation}
The $\ket{\alpha n}$ states are called the Floquet states and form an orthonormal basis. The time-evolution operator $U(t,t_0)$ can be written as
\begin{equation}\label{time_evolution}
    U_{\beta\alpha}(t,t_0)=\sum_{n=-\infty}^{\infty} \bra{\beta n}\mathrm{e}^{-iH_F(t-t_0)} \ket{\alpha 0}\mathrm{e}^{in\delta\omega t}.
\end{equation}

This $H_F$ Hamiltonian contains in its definition the $\delta\omega$ parameter, so it is not useful for calculations in this form. It was shown that the $H_F$ can be decomposed into block-diagonal form containing $H_{F,0}$, $H_{F,p_1}$, $H_{F,p_2}$ \ldots blocks \cite{ho1983semiclassical}. The $H_{F,0}$ block is defined through the subspace $G_0$ spanned by the basis set $\{\ket{\alpha m}\}$, where $m$ is an integer, which can be represented as $nN+\bar n\bar N$, $n$ and $\bar n$ are arbitrary integers. The other blocks are defined in a similar way, the subspaces $G_p$ are spanned by $\{\ket{\alpha,p+m}\}$, where $p$ is an integer, which cannot be written as $nN+\bar n\bar N$. In every subspace we can define the Floquet block if we relabel the $\ket{\alpha n}$ state as $\ket{\alpha pn\bar n}$,
\begin{eqnarray}
   &&\bra{\alpha pn\bar n}H_{F,p}\ket{\beta p k\bar k}=H_{\alpha\beta}^{(n-k,\bar n-\bar k)}\nonumber\\&+&(p\delta\omega+n\omega+\bar n\bar\omega)\delta_{\alpha\beta}\delta_{nk}\delta_{\bar n\bar k}.
\end{eqnarray}
The $H_{\alpha\beta}^{(n,\bar n)}$ Fourier component can be calculated if we use the Hamiltonian defined in Eq.~(\ref{general_Hamiltonian}),
\begin{eqnarray}
   H_{\alpha\beta}^{(n,\bar n)}&=&E_{\alpha}\delta_{\alpha\beta}\delta_{n,0}\delta_{\bar n,0}+\dfrac{V_{\alpha\beta}}{2}(\mathrm{e}^{i\phi}\delta_{n,1}+\mathrm{e}^{-i\phi}\delta_{n,-1})\delta_{\bar n,0}\nonumber\\&+& \dfrac{\bar V_{\alpha\beta}}{2}(\mathrm{e}^{i\bar\phi}\delta_{\bar n,1}+\mathrm{e}^{-i\bar\phi}\delta_{\bar n,-1})\delta_{n,0},
\end{eqnarray}
where $V_{\alpha\beta}=\bra{\alpha}V\ket{\beta}$, $\bar V_{\alpha\beta}=\bra{\alpha}\bar V\ket{\beta}$.
The time-evolution operator in Eq.~(\ref{time_evolution}) contains the $\ket{\alpha 0}$ vector, so it is sufficient to consider only the $H_{F,0}$ block of the $H_F$ Hamiltonian. We can denote the $\ket{\beta p k\bar k}$ as $\ket{\beta k \bar k}$, because we know that $p=0$. The time-evolution operator becomes
\begin{eqnarray}
    U_{\beta\alpha}(t,t_0)=&&\sum_{n=-\infty}^{\infty} \sum_{\bar n=-\infty}^{\infty} \bra{\beta n\bar n}\mathrm{e}^{-iH_{F,0}(t-t_0)}\ket{\alpha 00} \nonumber  \\ &  & \times\mathrm{e}^{i(n\omega+\bar n\bar\omega)t}.
\end{eqnarray}
The $H_{F,0}$ Hamiltonian describes the transitions as the Hamiltonian defined in Eq.~(\ref{general_Hamiltonian}) does, but it is time-independent. For convenience we denote the $H_{F,0}$ Hamiltonian as $\mathcal{H_F}$. For a more detailed investigation of the structure of the Hamiltonian see Ref. \onlinecite{ho1983semiclassical}.

\section{Calculation of bichromatic ESR}

\label{app_Schrieffer_Wolff}
In weak-driving limit, when $\left\{B_\textrm{ac},\bar B_\textrm{ac}\right\} \ll B$ is fulfilled,  a perturbative description of the multi-photon transitions is possible. Using time-independent Schrieffer-Wolff transformation \cite{schrieffer1966relation}, also known as quasidegenerate perturbation theory \cite{winkler2003quasi} we can reduce the infinite-dimensional Floquet Hamiltonian, $\mathcal{H_F}$, to an effective $2\times 2$ Hamiltonian which describes the bichromatic transition up to second order in the ac magnetic fields. The $V\cos{\omega t}$ and $\bar{V}\cos{\bar\omega t}$ driving terms of the Hamiltonian described in Eq.~(\ref{general_Hamiltonian}) are considered perturbations.  

The set of eigenfunctions of the Floquet Hamiltonian can be divided into weakly interacting subsets A and B, and we are only interested in the set A, which will describe the two-photon transition. The quasi-degenerate perturbation theory is a time-independent unitary transformation which transforms the Hamiltonian into a new one, which consists of two blocks, one corresponding to the set A, the other one to the set B. These two blocks are independent from each other, meaning that the transformed Hamiltonian is block-diagonal up to a given order in the perturbation. 

Only a finite part of the Hamiltonian gives contribution up to a given order in the perturbation. 
The matrix elements of the $\mathcal{H_F}$ Hamiltonian: 

\begin{widetext}
\begin{equation}\label{floquet_def}
    \scalebox{0.968}{$\bra{\alpha n\bar n}\mathcal{H_F}\ket{\beta k \bar k}=\left(E_\alpha+n\omega+\bar n\bar\omega\right)\delta_{\alpha\beta}\delta_{nk}\delta_{\bar n \bar k}+\dfrac{V_{\alpha\beta}}{2}\left(\mathrm{e}^{i\phi}\delta_{n-k,1}+\mathrm{e}^{-i\phi}\delta_{n-k,-1}\right)\delta_{\bar n \bar k}+\dfrac{\bar V_{\alpha\beta}}{2}\left(\mathrm{e}^{i\bar\phi}\delta_{\bar n-\bar k,1}+\mathrm{e}^{-i\bar\phi}\delta_{\bar n - \bar k,-1}\right)\delta_{nk}.$}
\end{equation}
\end{widetext}
For convenience we denote the $\alpha=0, n=0, \bar n=0, \beta=0, k=0,  \bar k=0$ element of the $\mathcal{H_F}$ as $\mathcal{H_F}_{00}$. The $2\times2$ effective Hamiltonian that will describe the two-photon transition consists of elements $\mathcal{H_F}_{00}$, $\mathcal{H_F}_{01}$, $\mathcal{H_F}_{10}$, and $\mathcal{H_F}_{11}$. The second-order correction: 
\begin{equation}\label{Quasi_2nd_order}
    \mathcal{H_F}_{mm'}^{[2]}=\dfrac{1}{2} \sum_{l\neq m,m'} \mathcal{H_F}_{ml}\mathcal{H_F}_{lm'}\left(\dfrac{1}{E_m-E_l}+\dfrac{1}{E_{m'}-E_l}\right),
\end{equation}
where $E_m$, $E_l$, $E_{l'}$ are diagonal elements of the Floquet Hamiltonian.  
We can get the Rabi frequency through calculating the second order correction of the $\mathcal{H_F}_{01}$ off-diagonal element, because $\left| \mathcal{H_F}_{01}^{[2]}\right|=\dfrac{\Omega_{\mathrm{Rabi},bi}}{2}$.  Using Eq.~(\ref{Quasi_2nd_order}) and the approximate condition of the resonance, $\mathcal{H_F}_{00}=\mathcal{H_F}_{11}$, we can calculate the $\mathcal{H_F}_{01}^{[2]}$, because only the $l=-2$ and $l=3$ cases give contribution to the sum.

If we take the Hamiltonian of the ESR described in Eq.~(\ref{ESR_Hamiltonian}), we can write the $8\times8$ part of the $\mathcal{H_F}$, which is sufficient to calculate the Rabi frequency,
\newcommand\myeq{\mkern1.5mu{=}\mkern1.5mu}
\usetikzlibrary{matrix}
\usetikzlibrary{calc,fit}
\tikzset{%
  highlight1/.style={rectangle,color=blue!,fill=blue!15,draw,fill opacity=0.3,thick,inner sep=0pt}
}
\tikzset{%
  highlight2/.style={rectangle,color=gray!,fill=gray!15,draw,fill opacity=0.3,thick,inner sep=0pt}
}

\begin{widetext}
\begin{equation} \tiny
\setlength{\arraycolsep}{2.5pt}
\medmuskip = 1mu  \mathcal{H_{F}}=
\begin{tikzpicture}[baseline=(m.center)]
  \filldraw[fill=blue!15, fill opacity=0.3,thick,draw=blue] (0.9,0.3) -- (-2.5,0.3) --(-2.5,-1.2)-- (0.9,-1.2)--(0.9,0.3);
   \filldraw[fill=gray!15, fill opacity=0.3,thick,draw=gray](-8.3,0.3) -- (-8.3,2.85)--(-2.5,2.85)--(-2.5,0.3)--(-8.3,0.3);
   \filldraw[fill=gray!15, fill opacity=0.3,thick,draw=gray](-8.3,-1.2) -- (-8.3,-3.75)--(-2.5,-3.75)--(-2.5,-1.2)--(-8.3,-1.2);
   \filldraw[fill=gray!15, fill opacity=0.3,thick,draw=gray](6.32,-1.2) -- (6.32,-3.75)--(0.9,-3.75)--(0.9,-1.2)--(6.32,-1.2);
   \filldraw[fill=gray!15, fill opacity=0.3,thick,draw=gray](6.32,2.85) -- (6.32,0.3)--(0.9,0.3)--(0.9,2.85)--(6.32,2.85);

    \matrix (m) [matrix of math nodes, left delimiter={.}, right delimiter={.},
     row sep=1mm, nodes={minimum width=2em, minimum height=1em}] {
 {} & \beta \myeq 1&  \beta \myeq 0  & \beta \myeq 1  & \beta \myeq 0   & \beta \myeq 1  & \beta \myeq 0 & \beta \myeq 1 & \beta \myeq 0 & {} \\[-0.2cm]
  {} &  k\myeq 1&  k \myeq 1  & k \myeq 0  & k \myeq 0   & k \myeq 1  & k \myeq 1 & k \myeq 0 & k \myeq 0 & {}  \\[-0.2cm]
 {} & \bar k \myeq 0& \bar k \myeq 0  & \bar k \myeq 0  & \bar k \myeq 0   & \bar k \myeq 1  & \bar k \myeq 1 & \bar k \myeq 1 & \bar k \myeq 1 & {} \\[-0.2cm]
  {} & \downarrow &  \downarrow  & \downarrow  & \downarrow  & \downarrow  & \downarrow & \downarrow & \downarrow & {} \\ [-0.3cm]
  {} & \vdots &  \vdots  & \vdots  & \vdots  & \vdots & \vdots  & \vdots & \vdots & {} \\ [-0.2cm]
 \hdots  & -\dfrac{\omega_{\mathrm{split}}}{2}+\omega & 0 & -\dfrac{g\mu_BB_{\mathrm{ac}}}{4}\mathrm{e}^{i\phi} & 0 & 0 & \dfrac{g\mu_B\bar B_{\mathrm{ac}}}{4}\mathrm{e}^{-i\bar\phi} &  0 & 0 & \hspace{-2mm} \hdots & \hspace{-6mm}  \leftarrow & \hspace{-5mm}  \alpha \myeq 1, n\myeq 1, \bar n\myeq 0 \\
 \hdots  & 0 & \dfrac{\omega_{\mathrm{split}}}{2}+\omega & 0  & \dfrac{g\mu_B B_{\mathrm{ac}}}{4}\mathrm{e}^{i\phi} & \dfrac{g\mu_B\bar B_{\mathrm{ac}}}{4}\mathrm{e}^{-i\bar\phi} & 0 &  0  & 0 & \hspace{-2mm} \hdots & \hspace{-6mm} \leftarrow & \hspace{-5mm} \alpha \myeq 0, n \myeq 1, \bar n \myeq 0 \\
  \hdots & -\dfrac{g\mu_B B_{\mathrm{ac}}}{4}\mathrm{e}^{-i\phi} & 0 & -\dfrac{\omega_{\mathrm{split}}}{2} & 0 & 0 &  0 & 0 & \dfrac{g\mu_B\bar B_{\mathrm{ac}}}{4}\mathrm{e}^{-i\bar\phi} & \hspace{-2mm} \hdots & \hspace{-6mm} \leftarrow & \hspace{-5mm} \alpha \myeq 1, n \myeq 0, \bar n \myeq 0 \\
  \hdots  & 0 & \dfrac{g\mu_BB_{\mathrm{ac}}}{4}\mathrm{e}^{-i\phi} & 0 & \dfrac{\omega_{\mathrm{split}}}{2} & 0 & 0 & \dfrac{g\mu_B\bar B_{\mathrm{ac}}}{4}\mathrm{e}^{-i\bar\phi} & 0 & \hspace{-2mm} \hdots & \hspace{-6mm} \leftarrow & \hspace{-5mm} \alpha \myeq 0, n \myeq 0, \bar n \myeq 0 \\
 \hdots  & 0 & \dfrac{g\mu_B\bar B_{\mathrm{ac}}}{4}\mathrm{e}^{i\bar\phi} & 0 & 0 & -\dfrac{\omega_{\mathrm{split}}}{2}+\omega+\bar\omega & 0 & -\dfrac{g\mu_BB_{\mathrm{ac}}}{4}\mathrm{e}^{i\phi} & 0 & \hspace{-2mm} \hdots & \hspace{-6mm} \leftarrow & \hspace{-5mm} \alpha \myeq 1, n \myeq 1, \bar n \myeq 1 \\
 \hdots & \dfrac{g\mu_B\bar B_{\mathrm{ac}}}{4}\mathrm{e}^{i\bar\phi} & 0 & 0 & 0 & 0 & \dfrac{\omega_{\mathrm{split}}}{2}+\omega+\bar\omega & 0 & \dfrac{g\mu_BB_{\mathrm{ac}}}{4}\mathrm{e}^{i\phi} & \hspace{-2mm} \hdots & \hspace{-6mm} \leftarrow & \hspace{-5mm} \alpha \myeq 0, n \myeq 1, \bar n \myeq 1 \\
 \hdots & 0 & 0 & 0 & \dfrac{g\mu_B\bar B_{\mathrm{ac}}}{4}\mathrm{e}^{i\bar\phi} & -\dfrac{g\mu_BB_{\mathrm{ac}}}{4}\mathrm{e}^{-i\phi} & 0 & -\dfrac{\omega_{\mathrm{split}}}{2}+\bar\omega & 0 & \hspace{-2mm} \hdots & \hspace{-6mm} \leftarrow & \hspace{-5mm} \alpha \myeq 1, n \myeq 0, \bar n \myeq 1 \\
 \hdots & 0 & 0 & \dfrac{g\mu_B\bar B_{\mathrm{ac}}}{4}\mathrm{e}^{i\bar\phi} & 0 & 0 & \dfrac{g\mu_BB_{\mathrm{ac}}}{4}\mathrm{e}^{-i\phi} & 0 & \dfrac{\omega_{\mathrm{split}}}{2}+\bar\omega & \hspace{-2mm} \hdots & \hspace{-6mm} \leftarrow & \hspace{-5mm} \alpha \myeq 0, n \myeq 0, \bar n \myeq 1 \\[-0.2cm]
{} & \vdots & \vdots & \vdots & \vdots  & \vdots & \vdots  & \vdots & \vdots & {} & {} & {}  \\
      }; 
%      \node[highlight1, fit= (m-5-6.north east)(m-6-5.south west)] {};
%      \node[highlight2, fit=(m-6-7.south west) (m-10-10.south east)] {};
%      \node[highlight2, fit=(m-5-6.north east) (m-2-10.south east)] {};
%      \node[highlight2, fit=(m-2-3.south west) (m-5-5.north west)] {};
      %\node[highlight2, fit=(m-6-5.south west) (m-10-3.south west)] {};
\end{tikzpicture} 
\label{eq:Floq_esr}
\end{equation}
\end{widetext}
 The result of the calculation: 
\begin{equation}\label{ESR_Rabi}
    \Omega_\textrm{Rabi,bi}=\dfrac{\left(g\mu_B\right)^2B_{\mathrm{ac}}\bar B_{\mathrm{ac}}}{4\hbar^2\omega} .
\end{equation}

In order to calculate the Bloch-Siegert shift we have to calculate the second order corrections to the diagonal elements of the effective $2\times 2$ Hamiltonian, $\mathcal{H_F}_{00}^{[2]}, \mathcal{H_F}_{11}^{[2]}$. We could carry out this calculation using a well chosen $8\times 8$ matrix, just like we did before, but it is not necessary. For the first diagonal element the Eq.~(\ref{Quasi_2nd_order}) that describes the second order correction gets simplified to: 
\begin{equation}\label{quasi_2nd_diagonal}
    \mathcal{H_F}_{00}^{[2]}=\sum_{l\neq 0} \dfrac{\abs{\mathcal{H_F}_{0l}}^2}{E_0-E_l}.
\end{equation}

This sum can be evaluated without arranging the $\bra{\alpha n\bar n}\mathcal{H_F}\ket{\beta k \bar k}$ elements in a matrix. The $\mathcal{H_F}_{0l}$ is the $\bra{000}\mathcal{H_F}\ket{\beta k \bar k}$ element, the $l\neq 0$ condition means that the $\beta, k, \bar k$ indices cannot be zero at the same time, but we do not indicate this fact in the next equations. The $E_m$ diagonal element can be indexed in fact with three indices, 
\begin{equation}
    E_m=E_\alpha+n\omega+\bar n\bar\omega=E_{\alpha n \bar n}, 
\end{equation}
where $\alpha$ $\in$ $\{0,1\}$, $E_0=\dfrac{\omega_{\textrm{split}}}{2}$, $E_1=-\dfrac{\omega_{\textrm{split}}}{2}$.  
The sum reformulated
\begin{equation}
    \mathcal{H_F}_{00}^{[2]}=\sum_{\beta=0}^1\sum_{k=-\infty}^{\infty}\sum_{\bar k=-\infty}^{\infty} \dfrac{\abs{\bra{000}\mathcal{H_F}\ket{\beta k \bar k}}^2}{E_{000}-E_{\beta k \bar k}}.
\end{equation}
Using Eq.~(\ref{floquet_def}) the sum can be calculated, the result is 
\begin{equation}\label{HF00}
    \mathcal{H_F}_{00}^{[2]}=\dfrac{\left(g\mu_B\right)^2\bar B_{\mathrm{ac}}^2}{16}\left(\dfrac{1}{\omega}+\dfrac{1}{\omega+2\bar\omega}\right) . 
\end{equation}
The second-order correction to the other diagonal element can be calculated similarly,
\begin{equation}
    \mathcal{H_F}_{11}^{[2]}=\sum_{\beta=0}^1\sum_{k=-\infty}^{\infty}\sum_{\bar k=-\infty}^{\infty} \dfrac{\abs{\bra{111}\mathcal{H_F}\ket{\beta k \bar k}}^2}{E_{111}-E_{\beta k \bar k}}.
\end{equation}
This yields
\begin{equation}\label{HF11}
    \mathcal{H_F}_{11}^{[2]}=-\dfrac{\left(g\mu_B\right)^2\bar B_{\mathrm{ac}}^2}{16}\left(\dfrac{1}{\omega}+\dfrac{1}{\omega+2\bar\omega}\right) .
\end{equation}
The effective $2\times 2$ Hamiltonian that describes the transition, $H_{\mathrm{eff}}$,
\begin{equation}\label{H_eff}
    H_{\mathrm{eff}}=\begin{pmatrix}
    \dfrac{\omega_{\textrm{split}}}{2}+\mathcal{H_F}_{00}^{[2]} & -\dfrac{\Omega_{\textrm{Rabi,bi}}}{2}\mathrm{e}^{-i(\phi+\bar\phi)} \\[6pt]
    -\dfrac{\Omega_{\textrm{Rabi,bi}}}{2}\mathrm{e}^{i(\phi+\bar\phi)} & -\dfrac{\omega_{\textrm{split}}}{2}+\omega+\bar\omega+\mathcal{H_F}_{11}^{[2]}
    \end{pmatrix} . 
\end{equation}
\par Note that the off-diagonal element of the effective Hamiltonian has a complex phase, it means that in the corotating frame the qubit state vector rotates around an axis that lies in the $x-y$ plane and forms an angle of ($\phi+\bar\phi$) with the $x$-axis. The frequency of the rotation is the Rabi frequency calculated in Eq.~(\ref{ESR_Rabi}). 
\par The resonance condition is that the two diagonal elements are equal, 
\begin{equation}
    \omega+\bar\omega=\omega_{\textrm{split}}+\mathcal{H_F}_{00}^{[2]}-\mathcal{H_F}_{11}^{[2]}.
\end{equation}
The term next to the $\omega_{\textrm{split}}$ is the drive-dependent shift of the resonance, the Bloch-Siegert shift. Using Eqs.~(\ref{HF00}), (\ref{HF11}) and the condition $\omega+\bar\omega\approx \omega_{\textrm{split}}$,  we get the  Bloch-Siegert frequency Eq. (\ref{ESR_BS}).

\par The calculation in the case of charge and flopping-mode spin qubit is similar, the only additional step is a basis transformation into energy basis of the undriven system before applying the Floquet theory. The basis transformation of the charge qubit Hamiltonian can be carried out exactly, while the transformation of the flopping-mode spin qubit Hamiltonian can be handled perturbatively, where the term describing the spin-orbit interaction and the g-factor antisymmetry are considered perturbations, so a hierarchy between the parameter holds as Eq. (\ref{condition}) suggests. 

\bibliography{paper}
\end{document}